\pdfoutput=1
\documentclass[aps,pre,floatfix,reprint,superscriptaddress,natbib]{revtex4-2} 
\def\fps@figure{!t}
\makeatother

\usepackage{amsmath,amssymb,amsthm,amsbsy,amsfonts,eurosym,latexsym,float,CJKutf8,comment}
\usepackage{xcolor,graphicx,subfigure,placeins,epstopdf}
\usepackage{tabularx,booktabs,multirow,array}

\definecolor{lgolred}{RGB}{221,20,8} 
\definecolor{lgolblue}{RGB}{26,72,157}
\definecolor{lgolgreen}{RGB}{23,146,50}

\usepackage{hyperref}

\hypersetup{
    pdfinfo={
		Title={Deterministic scale-invariant dynamics in a logistic Game-of-Life model},
		Author={Hakan Akg{\"u}n, Yan Xianquan, Tamer Ta\c{s}kıran, Muhamet Ibrahimi, Ching Hua Lee, Seymur Jahangirov},
	},
    colorlinks=true,
    linkcolor=blue,
    citecolor=blue,
    filecolor=magenta,
    urlcolor=blue,
}

\DeclareMathAlphabet\mathbfcal{OMS}{cmsy}{b}{n}

\DeclareMathOperator{\cO}{\mathcal{O}}
\DeclareMathOperator*{\argmin}{arg\,min}

\DeclareMathOperator{\erfc}{erfc}

\newcommand{\abs}[1]{\left|#1\right|}

\newcommand{\expect}[1]{\langle{#1}\rangle}

\newcommand{\cS}{\mathcal{S}}
\newcommand{\mbf}[1]{\mathbf{#1}}

\newcommand{\mrm}[1]{\mathrm{#1}}

\newcommand{\Eref}[1]{Eq.~(\ref{#1})}

\usepackage{comment}
\usepackage[english]{babel} 
\usepackage{float}
\usepackage{graphicx}
\usepackage{multirow} 
\usepackage{CJKutf8}
\usepackage{placeins}

\newcolumntype{P}[1]{>{\centering\arraybackslash}p{#1}}
\newcolumntype{M}[1]{>{\centering\arraybackslash}m{#1}}
 
\makeatletter
\newcommand{\beginsupplement}{%
  \setcounter{section}{0}%
  \setcounter{subsection}{0}%
  \setcounter{figure}{0}%
  \setcounter{table}{0}%

  \renewcommand{\thesection}{\arabic{section}}%

  \renewcommand{\thefigure}{S\arabic{figure}}%
  \renewcommand{\thetable}{S\arabic{table}}%

  \renewcommand{\@seccntformat}[1]{%
    \def\@tempa{##1}\def\@tempb{section}%
    \ifx\@tempa\@tempb
      SUPPLEMENTARY NOTE~\thesection:\quad
    \else
      \csname the##1\endcsname\quad
    \fi
  }%

  \renewcommand{\fnum@figure}{Supplementary Figure~\thefigure}%
  \renewcommand{\fnum@table}{Supplementary Table~\thetable}%
}
\makeatother

\newcommand{\lA}{\lambda_{\mathrm{A}}}
\newcommand{\lP}{\lambda_{\mathrm{P}}}

\newcommand{\dc}{d_{\mathrm{c}}}
\newcommand{\df}{d_{\mathrm{f}}}

\newcommand{\sigf}{\sigma_{\mathrm{f}}}

\newcommand{\RW}[1]{R_{\mathrm{W}}^{(\mathrm{#1})}} 
 
\newcommand{\SupNoteref}[1]{Supplementary Note~\ref{#1}}

\newcommand{\MainFigref}[1]{Figure~\ref{#1}}

\newcommand{\MainTableref}[1]{Table~\ref{#1}}

\newcommand{\SupNoteFigref}[1]{Supplementary Figure~\ref{#1}}
\newcommand{\SupNoteTableref}[1]{Supplementary Table~\ref{#1}}

\renewcommand{\thesection}{\arabic{section}}
\renewcommand{\thetable}{\arabic{table}}

\begin{document}

\begin{CJK}{UTF8}{gbsn}
\title{Deterministic scale-invariant dynamics in a logistic Game-of-Life model}

\author{Hakan Akg{\"u}n}\email{hakan.akgun@u.nus.edu} 
\affiliation{Department of Physics, National University of Singapore, Singapore 117551} 

\author{Xianquan Yan (颜显权)} 
\affiliation{Department of Physics, National University of Singapore, Singapore 117551}
\affiliation{Department of Computer Science, National University of Singapore, Singapore 117417}

\author{Tamer Ta\c{s}kıran} 
\affiliation{UNAM, Institute of Materials Science and Nanotechnology, Bilkent University, Ankara 06800, Turkey} 

\author{Muhamet Ibrahimi} 
\affiliation{Laboratory of Artificial and Natural Evolution (LANE), Department of Genetics and Evolution, \\
University of Geneva, 1211 Geneva, Switzerland}

\author{Ching Hua Lee}\email{phylch@nus.edu.sg}
\affiliation{Department of Physics, National University of Singapore, Singapore 117551}

\author{Seymur Jahangirov}\email{seymur@unam.bilkent.edu.tr}
\affiliation{UNAM, Institute of Materials Science and Nanotechnology, Bilkent University, Ankara 06800, Turkey}

\date{\today}

\begin{abstract}

Scale invariance is a hallmark of criticality in complex dynamical systems. While random external inputs or tunable stochastic interactions are typically required to produce critical behavior, it remains unclear whether scale-invariant dynamics can emerge from purely deterministic interactions. Here, we address this question by studying the asymptotic dynamics of the \textit{logistic} Game of Life (GOL), a deterministic-parameter extension of Conway’s GOL. In this system, we identify three distinct asymptotic phases separated by two fundamentally different critical points. The first critical point, associated with an unusual form of self-organized criticality, separates a sparse-static phase from a sparse-dynamic phase. The second critical point corresponds to a deterministic percolation transition between the sparse-dynamic phase and a third, dense-dynamic phase. In addition, we observe power-law cluster size distributions with unconventional critical exponents not found in standard equilibrium systems. Overall, our work paves the way for studying emergent scale invariance in purely deterministic systems.

\end{abstract}

\maketitle
\end{CJK}

\section{Introduction}\label{sec:intro} 

Scale invariance is a hallmark of critical behavior in dynamical systems~\cite{odor2004universality, lee2017generalized, arouca2020, rafi_ul_islam2022,liu2024non,Parisi2002,Creswick_1997,Talapov_1996}. In particular, spatially extended systems driven by local interactions exhibit scale-invariant dynamics by organizing in clusters with no characteristic size and/or duration. Typically, such behavior emerges either from the intrinsic characteristics of interactions -- i.e., known as self-organized criticality (SOC) -- or from an external tuning parameter that modulates the strength of interactions -- i.e., parameter-driven criticality ~\cite{Khaluf2017}. The former has been identified in abelian sandpile~\cite{bak1987self,bak1988self}, forest-fire~\cite{Drossel1992}, and earthquake~\cite{Olami1992} cellular automaton models, which, although driven by deterministic toppling dynamics, still depend on stochastic grain addition.  In contrast, the latter is realized in systems that undergo percolation transitions, where typically a probabilistic control parameter—such as the site/bond occupation probability \(p\)—is tuned through a threshold, yielding universal scaling laws~\cite{yang2024,stauffer1979,vardi1999,Stauffer1994,sahimi1994applications,Sheinman2015,Federbush2021,Hu2016}. This raises the question of whether scale-invariant dynamics can originate solely from deterministic interactions, without any stochastic inputs or external noise.

To this end, multiple studies have examined the emergence of criticality in deterministic systems from various perspectives. For example, invasion~\cite{peng1990} and bootstrap percolation~\cite{vardi1999, bootstrap2021}, random walks~\cite{Terifmode2007,FREUND1992,BUNIMOVICH2004,SANTOS2007,Lima2001,Boyer2009}, fractal networks~\cite{clerc1985,Hamburger96,Limat1988} are shown to have analogous forms in partially or fully deterministic settings. Moreover, a deterministic ansatz for fractal-like critical snapshots~\cite{lee2016} has been proposed. It has also been shown that kinetic constraints to deterministic spin systems may bring directed percolations \cite{deger2022,Deger2022Arrest}. Lastly, other studies have drawn parallels between transitions in deterministic coupled map lattices~\cite{CHATE1988,GRASSBERGER1991,CUCHE1997} and percolation phenomena. Despite these works, clear evidence of purely deterministic percolation and phase transitions in 2D models remains elusive and has not been investigated through explicit cluster analyses. Here, we find affirmative signs of such critical behavior and point toward its potential occurrence in real-world systems.

One of the simplest deterministic systems that has been used for studying critical dynamic behavior is Conway's Game of Life (GOL).
This cellular automaton --defined by local parallel interactions (i.e., rules) among binary states in a square lattice of sites~\cite{gardner1970game}-- has often been a starting point for studying phenomena related to artificial life~\cite{Turney2021,Packard2024}, ecology~\cite{Faux2023}, and
self-organization~\cite{Rainwater2024,adamatzky2010,Ibrahimi2019,Ibrahimi2022}. Importantly, the underlying interactions of this system have also been `probed' for their capacity to exhibit self-organized criticality~\cite{bak1989self,Alstrom1994}, or to undergo critical phase transitions~\cite{Vieira2021,Nordfalk1996,Monetti1995,Huang2003,reia2014conway}. Regarding the latter, prior studies have extended GOL's dynamics with control parameters that, by incrementally modifying the rules away from the original system~\cite{blok1999asynchronous,Vieira2021,schulman1978statistical,Nordfalk1996,Monetti1995}, suggest that Conway's GOL rules are strongly associated with scale-invariant dynamics. However, while such variations employ stochastic components, this system has never been investigated in the context of deterministic critical behavior.

In this paper, we analyze the scale-invariant dynamics that emerge due to the phase transitions occurring in the logistic GOL~\cite{Ibrahimi2019}: a deterministic extension inspired by the logistic map~\cite{May1976}, where a control parameter changes the rate of update of sites by expanding the initially binary state space into a Cantor set. 
Such a self-similar state space consequently allows for `slower' interactions among sites, giving rise to dynamical trajectories different from Conway's GOL. As we tune the control parameter, we observe that the asymptotic dynamics of this system change from a sparse-static (I) phase (initially similar to Conway's GOL) to a sparse-dynamic (II), and then to a dense-dynamic (III) phase. We identify the points that separate these three distinct dynamical regimes numerically and study their critical properties by in-depth analyses of cluster dynamics. We find that the critical point separating phases I and II defines the boundary of a peculiar form of self-organized criticality in the sparse-dynamic phase, where quiescent clusters surrounded by active sites follow a power-law distribution (Fisher exponent $\tau \simeq 2.9$). Moreover, detailed cluster analyses at the critical point separating phases II and III ($\tau \simeq 1.81$), indicate a deterministic percolation transition.

The main results of this paper are that we pinpoint a purely deterministic system --i.e., a system without noise or stochasticity in interactions--that displays scale-invariant dynamics in both of its typical forms, namely percolation and SOC. Such dynamics are related to the physics of critical phenomena in different ways. First, one of the critical points reveals a new class of SOC behavior, which lies beyond traditional stochastic and Abelian frameworks. Considering the widespread role of SOC in describing real-world phenomena~\cite{Markovi2014,Tadi2021}, our findings suggest a new dynamical pathway to ‘reaching’ self-organized criticality. Second, in our nonequilibrium system, the Fisher exponent at the percolation point is unconventional ($\tau < 2$) and cannot occur in standard equilibrium systems, as it would violate the hyperscaling constraints that apply in those cases.~\cite{Stauffer1994,sahimi1994applications} Interestingly, the same exponent has been observed only in no-enclave percolation~\cite{Sheinman2015}, which is known to originate from the backbone clusters of random percolation systems~\cite{Hu2016}. Our deterministic model demonstrates that radial anisotropy embedded in local update rules can generate the same anomalous exponent in regular clusters.

\begin{figure}[ht] 
\centering
\includegraphics[width=\linewidth]{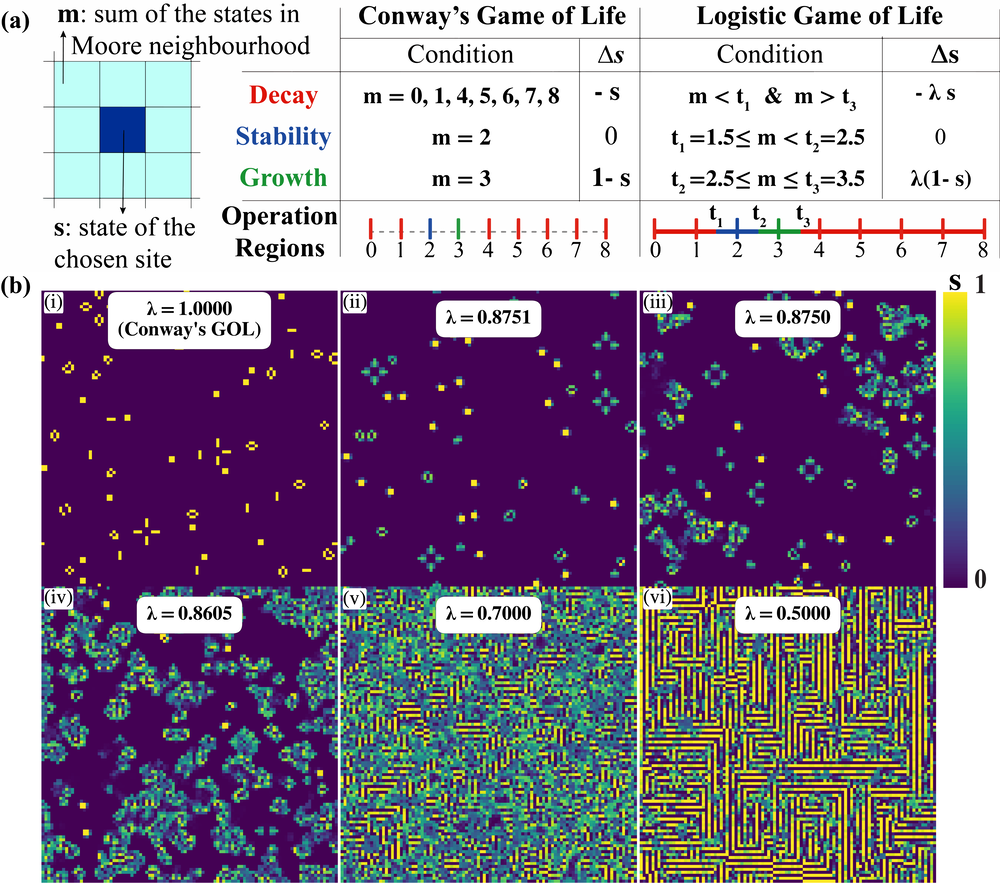}
\caption{\textbf{The logistic Game of Life.}
\textbf{(a)} Summary of the update rules of Conway’s Game of Life (middle) and the logistic Game of Life (right).
In the logistic Game of Life, each lattice site carries a continuous state $s\in[0,1]$, and the control parameter $\lambda$ sets the update strength.
For a site at time $t$ with state $s^t$, we compute the Moore sum $m^t$ and define the increment $\Delta s \equiv s^{t+1}-s^t$, which depends on three thresholds $t_1<t_2<t_3$ that partition the neighborhood-sum axis into stability, growth, and decay regimes.
Conway’s Game of Life is recovered in the discrete limit ($s\in\{0,1\}$) at $\lambda=1$ with the usual birth/survival rules.
\textbf{(b)} Illustrative snapshots of asymptotic configurations of the logistic Game of Life at representative values of the control parameter $\lambda$. The colorbar encodes the local site state $s$: colors toward dark purple indicate $s\simeq 0$, whereas colors toward bright yellow indicate $s\simeq 1$.}

\label{fig:model}
\end{figure}
\section{Results}
\subsection{The Logistic Game of Life}\label{sec:model}

The original Conway's GOL is defined on a square lattice of sites, where each site goes through the parallel updating scheme:
\begin{equation}
    s_j^{t+1} = s_j^t + \Delta s_j, \label{eq:gol}
\end{equation}
where $s_j^t \in \{0,1\}$ corresponds to the state of $j^{\text{th}}$ site at time point $t$. $\Delta s_j$ denotes the quantity to be added to update the state from $s_j^t$ to $s_j^{t+1}$, and is a function of $s_j^t$ itself and the sum of states in its Moore neighborhood $m_j^t$ (\MainFigref{fig:model}a, left panel), with $0\leq m_j^t \leq 8$. According to the finite-difference notation of Eq. \eqref{eq:gol}, a site in Conway's GOL can experience three possible updates: decay ($\Delta s_j = -s_j$ when $m_j<2$ or $m_j>3$), stability ($\Delta s_j = 0$ when $m_j=2$), or growth ($\Delta s_j = 1 - s_j$ when $m_j=3$) (\MainFigref{fig:model}a, middle).

On the other hand, the logistic GOL~\cite{Ibrahimi2019} stands as a prominent candidate for investigating deterministic scale-invariant dynamics in 2D systems. This system introduces a control parameter, $\lambda$, which tunes the update dynamics by rescaling the growth/decay rate of each site as $\Delta s_j^t \rightarrow \lambda \Delta s_j^t$, where $0 < \lambda \leq 1$. The case $\lambda = 1$ corresponds to the original limit of Conway's Game of Life (\MainFigref{fig:model}a, right). 

An important consequence of $\lambda$ in the logistic GOL is that the previously binary state space of the automaton expands into a Cantor set. To see this, one may define a simple representation that associates the three possible updates with discrete operators, respectively for decay (\textbf{D}), stability (\textbf{S}), and growth (\textbf{G}), such that:
\begin{equation}
    \textbf{D}s := (1-\lambda)s, \ \ \textbf{S}s := s, \ \ \textbf{G}s := (1-\lambda)s + \lambda.
\end{equation}
These discrete operators show how the state $s$ of a site may be updated, based on $\lambda$ and the nearest neighborhood (\MainFigref{fig:model}a, right). If we apply these operators to an initial set of $\{0,1\}$ once, they give rise to a larger set $\{0, 1-\lambda, \lambda, 1\}$. Again, applying operators to the new set gives rise to $\{0, (1-\lambda)^2, (1-\lambda)-(1-\lambda)^2, 1-\lambda, \lambda, \lambda + (1-\lambda)^2, \lambda, 1 - (1-\lambda)^2,  1\}$.  Repeating this recursively would lead to a $\lambda$ dependent Cantor set in the range $[0, 1]$. For later use, we define the order of each element in the Cantor set as the number of times a $\mbf{D}$ or $\mbf{G}$ operator has been applied to obtain it, starting from 0 or 1. For example, $\mbf{GG}\,0 = 1 - (1-\lambda)^2$  is a second-order Cantor value (see \SupNoteref{sec:LGOL_implementation} for details). 

A second consequence of $\lambda$ is that, due to the expanded state space, the space of neighborhood sums $m$ (which determine how sites are updated) is also expanded. In the logistic GOL, possible $m$ values span the range $[0,8]$ and comprise an eight-fold convolution of the Cantor set. To account for this, we assign two unit-length intervals centered at $m=2$ and $m=3$ as the neighborhood sum regions of stability and growth, respectively (\MainFigref{fig:model}b). We denote the limits of these intervals by $t_1 = 1.5, t_2 = 2.5,$ and $t_3 = 3.5$, such that sites get updated in the following fashion:
\begin{gather}
s^{t+1}_j =
\begin{cases}\label{eq:LGOL_rules}
    \mbf{S} s^{t}_j =  s^{t}_j & \text{if } t_1 \leq  m^{t}_j < t_2 \\
    \mbf{G} s^{t}_j =  (1- \lambda)s_j^t + \lambda & \text{if } t_2 \leq  m^t_j \leq t_3 \\
    \mbf{D} s_j^t =  (1-\lambda)s_j^t  & \text{ \ \ otherwise}
\end{cases}
\end{gather}

The rules of Conway's and logistic GOL are summarized in \MainFigref{fig:model}a, and snapshots of the asymptotic behavior of the logistic GOL at various $\lambda$ are displayed in \MainFigref{fig:model}b. As previously identified~\cite{Ibrahimi2019}, the dynamic and asymptotic behavior of logistic GOL for $0.875<\lambda<1$ is similar to Conway's GOL, where the system settles to a sparse inactive asymptotic state. Whereas for $\lambda \leq 0.875$, the system possesses active asymptotic states, which increasingly cover the system as $\lambda$ decreases (\MainFigref{fig:model}b). While previous work has discussed the asymptotic density around $\lambda = 0.875$ and the maze-like striped patterns at $\lambda < 0.7$~\cite{Ibrahimi2019}, no proper critical behavior has been identified. Here, we identify two points with distinct critical properties -- marking the boundaries between different asymptotic phases (\MainFigref{fig:model}b(iii)-(iv)\,) -- and characterize them through cluster analyses and power-law distributions. Importantly, the critical behaviors discussed here are independent of the initialization density: as long as the grid remains active, it converges to the same behavior regardless of the initial conditions (see \SupNoteref{sec:invariancetoinitialconds} for details). This distinguishes our model from others that rely on random initial conditions as effective control parameters, even though their update rules are deterministic.

To study the critical properties of the asymptotic dynamics, we perform simulations of the logistic GOL where the state space is truncated up to the $10^{\text{th}}$ order of the Cantor set. In other words, during simulations, any state with a higher-order Cantor value is `lumped' into the nearest Cantor value of order $\leq 10$ (see \SupNoteref{sec:LGOL_implementation} for implementation). Although this order is arbitrary, we note that the asymptotic behavior of the system remains unchanged if the Cantor set is truncated at higher orders.

\subsection{Signatures of Critical Behavior}\label{sec:critical} 

In this section, we study the asymptotic behavior of the logistic GOL, which exhibits remarkable changes as the control parameter $\lambda$ `drifts' the system away from Conway's GOL (see the different panels in \MainFigref{fig:model}b). To investigate whether such changes in the asymptotic behavior are related to critical phenomena, we define three quantities that characterize the system. 

First, we define an activity ($A^t$) order parameter of the following form:
\begin{equation}
    A^t := 1 - \frac{1}{N^2}\sum_{j} \delta_{s_j^t,\,s_j^{t-\bar{t}}}\label{eq:activity}
\end{equation}
where $N$ denotes the length of the square lattice, $\delta_{i,j}$ denotes the Kronecker delta, and the sum is over all sites. $A^t$ is thus defined to denote the fraction of cells that change states at time step $t$ after a time lag interval $\bar{t}$, serving as a measure of the lattice's autocorrelation. In the following, we set $\bar{t} = 60$ to exclude asymptotic-state oscillators with periods that are divisors of 60 \cite{adamatzky2010}. We then average the activity over time and ensemble to obtain $\langle A \rangle$. Thus, $\langle A \rangle = 1$ indicates that there is no autocorrelation between the states and their time-lagged counterparts (as expected from a fully active state), whereas $\langle A \rangle = 0$ reflects perfect autocorrelation, i.e. the grid's time-lagged version is identical to the current state.

Second, we use the definition of Eq.~\eqref{eq:activity} to characterize the spatio-temporal variation of activity through the susceptibility, defined as the fluctuation of the order parameter:
\begin{equation}
       \langle\chi\rangle := \langle A^2 \rangle - \langle A \rangle^2. 
\end{equation}
Analogous to magnetic systems, the susceptibility measures how uniformly the activity is distributed across the lattice. A system comprising only a few localized active sites is characterized by a high susceptibility, whereas a uniform distribution of active sites leads to a vanishing susceptibility.

Third, a cluster is a set of equal-state sites connected via their nearest neighbors (up, down, left, and right cells). Its size, $\cS^t$, is defined as the number of sites it contains, and at time $t$, clusters are ranked by size as $\cS_1^t \ge \cS_2^t \ge \cS_3^t \ge \dots$, with the index $i$ indicating the size rank. The cluster sizes are then averaged over time and ensemble to obtain $\langle \cS_i \rangle$.

\begin{figure}
\centering
\includegraphics[width=\linewidth]{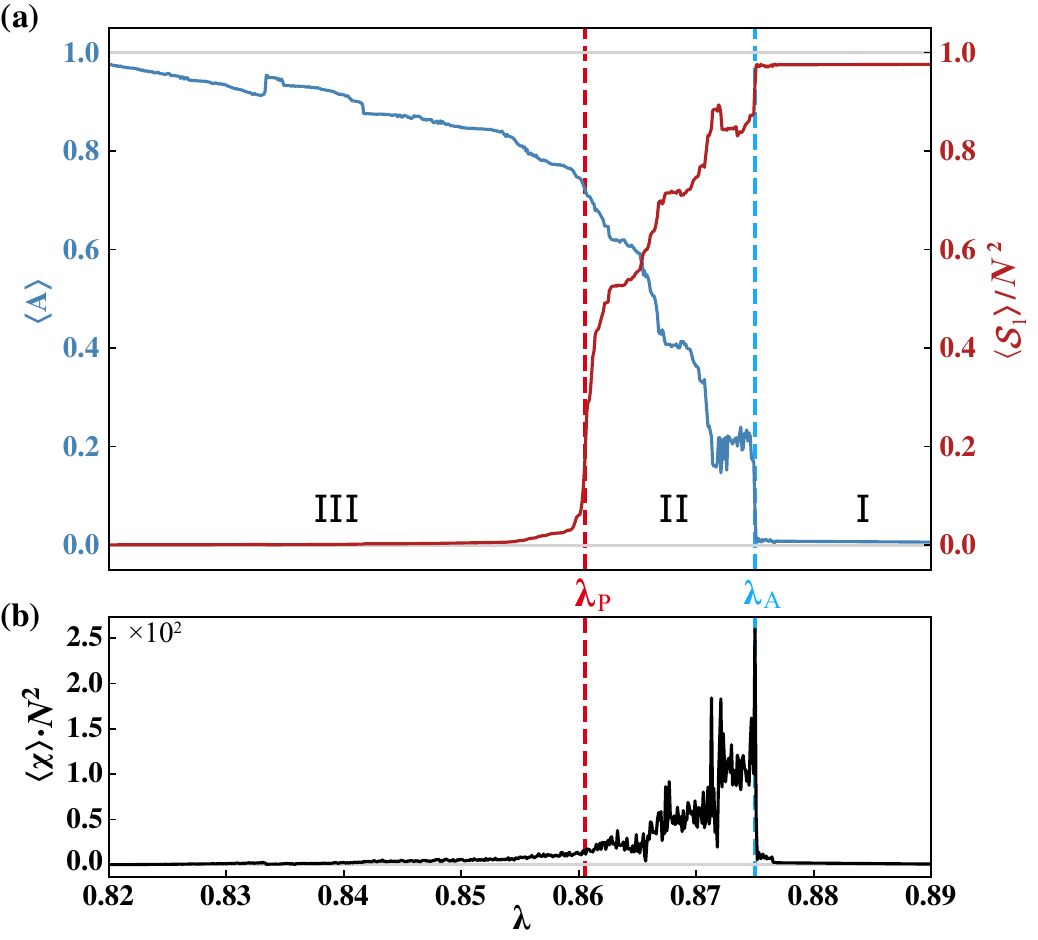}
    \caption{\textbf{Three distinct asymptotic phases in the logistic GOL separated by two critical points.} \textbf{(a)} Asymptotic average activity $\expect{A}$ (solid blue) and the size of the largest cluster $\expect{\cS_1}/N^2$ (solid red) computed against $\lambda$. The data indicate two critical points: (i) $\lA=0.875$ (blue dashed line), the boundary between a sparse-static (I) and sparse-dynamic (II) asymptotic phase; (ii) $\lP=0.86055$ (red dashed line), where fragmentation of the largest cluster defines the boundary between phase II and dense-dynamic (III) phase.  \textbf{(b)} The susceptibility of activity $\expect{\chi}$, plotted against $\lambda$, reaches its maximum at $\lA$.}
\label{fig:actcluster}
\end{figure}
 
In \MainFigref{fig:actcluster}, we report the numerically computed asymptotic quantities of $\langle A \rangle, \langle\chi\rangle$ and $\langle \cS_1 \rangle$ for the logistic GOL, where $\lambda$ is a control parameter. We focus on the parameter region $0.8<\lambda<0.9$, where we notice signs of critical behavior at $\lP = 0.86055$ and $\lA = 0.875$, and identify three distinct asymptotic phases of the system.

It is important to note that, while $\lA=0.875$ appears to be independent of the system size (see \SupNoteref{sec:SizeEffectsOnA} for details), the effective percolation point $\lP(N)=0.86055$ shifts slightly to the right as $N$ increases, with gradually diminishing corrections (see \SupNoteref{sec:Second_largest_divergence} for details). The thermodynamic percolation point $\lP(N\to\infty)=0.86134$ is identified from the wrapping probability. Since there are two closely related notions here, we state our convention explicitly: whenever we use $\lP$ without further qualification, we refer to the effective percolation point found at our experimental grid size $N=1024$, i.e., $\lP(N=1024)$, and we always refer to the thermodynamic percolation point as $\lP(N\to\infty)$.
In the following, we discuss in detail the asymptotic dynamics of the system as $\lambda$ changes. 

\subsubsection{$\lA$: Phase transition in asymptotic activity}

First, we discuss how the average asymptotic activity $\langle A \rangle$ changes in the logistic GOL as $\lambda$ is tuned down (blue data points in \MainFigref{fig:actcluster}a). 
We observe that $\langle A \rangle \simeq 0$ for $\lambda > \lA = 0.875 $, indicating that, in this parameter range, the logistic GOL comprises inactive asymptotic states. 
Indeed, similarly to Conway's GOL, the system settles to a sparse-static phase, i.e. a phase that is mostly populated by the vacuum background of quiescent states, and sparsely populated by stable blocks and periodic oscillators ( panels (i) and (ii) in \MainFigref{fig:model}b ). 
At $\lambda=\lA$ we observe a sudden increase in activity, which indicates that the asymptotic dynamics becomes fundamentally different. In the $\lambda \leq \lA$ range, the system does not settle into a static phase (panel (iii) in \MainFigref{fig:model}b), but rather it persists indefinitely in the thermodynamic limit. Incidentally, this reflects the recovery of ergodicity, where the system no longer converges to a single final configuration but visits all the possible configurations. 

Moreover, we find that the susceptibility $\langle\chi\rangle$ (\MainFigref{fig:actcluster}b) reaches its maximum at $\lambda=\lA$ (blue dashed line in \MainFigref{fig:actcluster}). The sudden jump in $\langle A \rangle$ and maximal $\langle\chi\rangle$ suggest that $\lA$ is the critical point which marks the static-dynamic transition in the asymptotic behavior of the system.

Besides defining the transition point between static and dynamic phases in the logistic GOL, the asymptotic susceptibility $\langle\chi\rangle$ provides additional insights on the nature of this transition. The fact that $\langle\chi\rangle$ increases sharply from zero to a maximum as $\lambda$ hits $\lA$ (\MainFigref{fig:actcluster}b), indicates that the asymptotic activity at the transition point is initially localized in a very low number of sites, and that the lattice is otherwise similar to the static phase in $\lA<\lambda\leq 1$ (panel (iii) in \MainFigref{fig:model}b). Moreover, the drop of $\langle\chi\rangle$ as $\lambda$ decreases below $\lA$, indicates that the activity becomes increasingly more spread in space, until the lattice becomes homogeneously active and $\langle\chi\rangle$ hence vanishes (panels (iv) and (v) in \MainFigref{fig:model}b and \MainFigref{fig:actcluster}b). 
 
\subsubsection{$\lP$: Phase transition in asymptotic cluster size}

Next, we identify a third asymptotic phase that emerges as $\lambda$ is tuned down even further. In particular, we investigate how the size of the largest cluster $\expect{\cS_1} /N^2$ -- i.e. corresponding to the vacuum cluster of quiescent states in the lattice -- changes with $\lambda$. The vacuum cluster (red in Figure~\ref{fig:actcluster}a) covers most of the grid when  $\lambda>\lA$. As $\lambda$ is tuned down below $\lA$, the size of the largest vacuum cluster drops, approximately following the inverse pattern of $\langle A\rangle$. However, as $\lambda$ decreases further, the behavior of $\expect{\cS_1}/N^2$ becomes remarkably different as compared to $\langle A\rangle$. The largest cluster of quiescent states experiences a sharp decrease, where the strongest drop occurs at $\lP \approx 0.86055$, defining another critical point.  

This sharp decrease in the size of the largest cluster (red dashed line in \MainFigref{fig:actcluster}) is important because it indicates a transition from an asymptotic dynamical phase with the vacuum cluster spanning the lattice, to a dynamical phase where there is no spanning cluster, and which is reminiscent of a percolation transition studied in the next section. We additionally note that the decrease in $\langle \chi \rangle$ (\MainFigref{fig:actcluster}b) as $\lambda$ goes below $\lP$ implies a more uniform activity within the lattice, and is another indicator of this third asymptotic phase.

\begin{table}[t]
\centering
\caption{\textbf{Operational transition neighborhoods at critical points.} The table above summarizes the neighborhood sums of the critical points at the operational thresholds \((t_1 = 1.5,\, t_2 = 2.5,\, t_3 = 3.5)\). The panel below shows neighborhoods undergoing transition, with unordered individual site values (as only the sum $m$ determines the operational region) highlighted around \(\lA = 0.875\), while the lower left panel illustrates the numerical evolution of these neighborhoods as \(\lambda\) varies between \(0 < \lambda < 1\). At the critical points, the polynomial neighborhoods switch regimes—G\(\leftrightarrow\)D and S\(\leftrightarrow\)D—corresponding to transitions in neighborhood sums \(t_3 \leftrightarrow t_1\) and \(t_2 \leftrightarrow t_1\), respectively, highlighting their role in the phase transition at \(\lA\).}
\label{tab:transitions-neighborhood-main}
\renewcommand{\arraystretch}{1.35}

\begin{tabular}{|c|c|l|}
\hline
$\lambda$ & Transition & Neighborhood \\ \hline

0.86055 & $\mathrm{G}\leftrightarrow\mathrm{D}$ &
\parbox[t]{0.62\columnwidth}{$
t_{3}\approx
\begin{aligned}[t]
&-3\lambda^{5}+16\lambda^{4}-34\lambda^{3}\\
&\quad +33\lambda^{2}-17\lambda+8
\end{aligned}
$} \\ \hline

0.86055 & $\mathrm{S}\leftrightarrow\mathrm{D}$ &
\parbox[t]{0.62\columnwidth}{$
t_{1}\approx
\begin{aligned}[t]
&-3\lambda^{5}+16\lambda^{4}-34\lambda^{3}\\
&\quad +33\lambda^{2}-17\lambda+6
\end{aligned}
$} \\ \hline

0.8750 & $\mathrm{G}\leftrightarrow\mathrm{D}$ &
\parbox[t]{0.62\columnwidth}{$t_{3}=4\lambda$} \\ \hline

0.8750 & $\mathrm{S}\leftrightarrow\mathrm{D}$ &
\parbox[t]{0.62\columnwidth}{$t_{1}=4(1-\lambda)+1$} \\ \hline

\multicolumn{3}{|c|}{%
\begin{minipage}{\columnwidth}
\vspace{1mm} 
\centering
\includegraphics[width=1\linewidth]{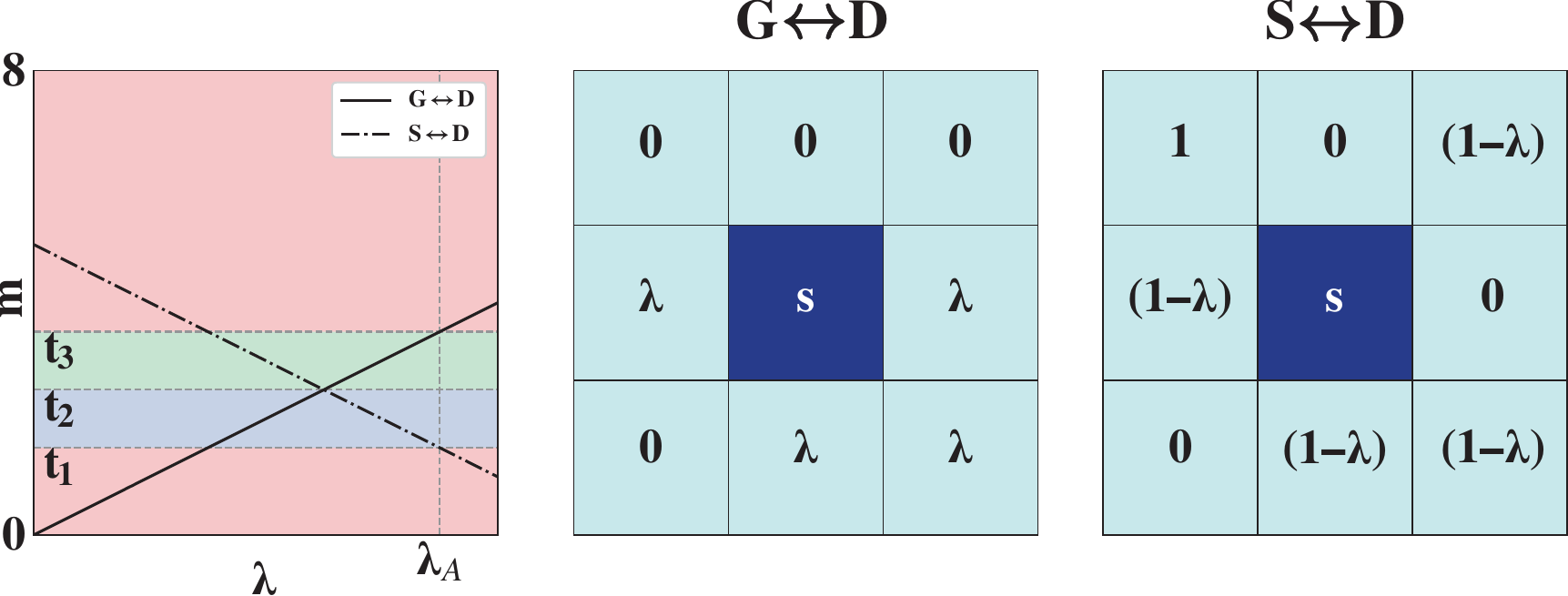}
\vspace{0.2mm} 
\end{minipage}
} \\ \hline

\end{tabular}
\end{table}

\subsubsection{Determination of phase transition points from GOL operational regions}

The increase in activity and the decrease of the vacuum cluster's size indicate that the average density of the system increases as $\lambda$ is tuned down (see also \MainFigref{fig:model}b). This occurs because, as $\lambda$ decreases, there are several neighborhood configurations which change their operation regions (\MainTableref{tab:transitions-neighborhood-main}). For example, a neighborhood $m_j$ consisting of $4\times \lambda$ sites and $4\times 0$ sites would `act' to decay the central site if $\lambda > \lA$ because $m_j=4\lambda > t_3=3.5$. However, for $\lambda\leq\lA$, then $m_j\leq t_3$, indicating that the central site will experience growth instead of decay. In a similar fashion, as $\lambda$ decreases, another neighborhood with $m_k = 1 + 4(1-\lambda)$ changes the operation region from decay to stability at $\lA$. In this case, the central site will decay when $\lambda>\lA$, as $m_k<t_1=1.5$; and it will remain stable when $\lambda\leq\lA$, as $m_k\geq t_1$. Note that there is a large set of neighborhood sums that changes operation regions as $\lambda$ is tuned down further, and it is these changes which alter the dynamics of the logistic GOL~\cite{Ibrahimi2019}. The main neighborhood sums that change operation regions at $\lambda=\lP$ are reported in \MainTableref{tab:transitions-neighborhood-main}.

While the transition at \(\lA\) reflects the influence of a fixed first-order Cantor-set polynomial Moore neighborhood, the fragmentation of the vacuum cluster near \(\lP\) emerges due to gradual neighborhood changes from higher-order polynomials. These gradual neighborhood transitions with changing \(\lambda\) govern the evolution of cluster shape, size, and scaling. Furthermore, since effective $\lP$ changes with the grid size $N$, the neighborhood associated with the \(\lP\) transition also changes with \(N\) (see \SupNoteref{sec:explicit_lambda_P_neighborhood} for details).

\begin{figure*}
\centering
\includegraphics[width=0.99\linewidth]{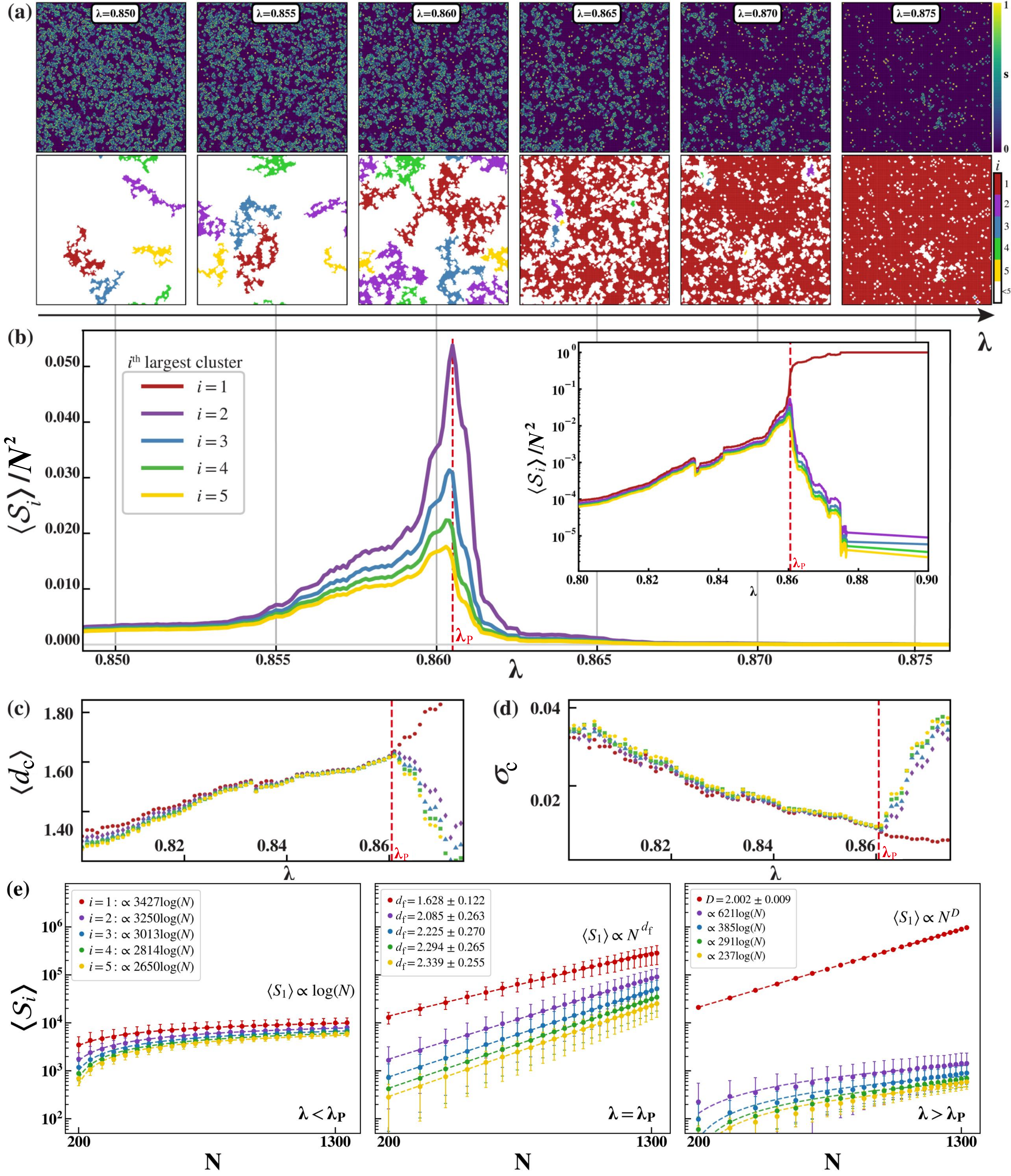}

\caption{\scriptsize 
\textbf{Deterministic cluster dynamics reveals a percolation transition in the logistic GOL.} \textbf{(a)} Top panels display snapshots of the asymptotic states of the logistic GOL at distinct $\lambda$ values in the range $[0.850,0.875]$ Bottom panels show the corresponding five largest clusters masked in different colors
(ranking in panel b), while the red dashed line marks $\lP$.
\textbf{(b)} Sizes of the largest clusters $\expect{\cS_2} \sim \expect{\cS_5}$ plotted against $\lambda$, where the index $i$ indicates the size rank of the cluster. The curves differ only by scaling when $\lambda<\lP$. The inset displays the logarithmic evolution of cluster sizes, with the largest zero-state cluster (dark red) percolating as \( \lambda \) increases.
The evolution of \textbf{(c)} capacity dimensions $\dc$ of the largest clusters and their \textbf{(d)} corresponding standard deviations $\sigma_c$ computed as functions of $\lambda$. \textbf{(e)} Scaling behavior of the largest cluster with lattice size ($N$) around \(\lP\). In the `subcritical' regime ($\lambda<\lP$, left), cluster sizes $ \langle \cS_i(N) \rangle $ follow a logarithmic trend. Around the critical point ($\lambda = \lP$, middle), the clusters scale as power laws, where the exponent of the largest cluster defines the fractal dimension $\df$. In the `supercritical' regime ($\lambda > \lP$, right), the largest cluster $ \langle \cS_1 \rangle $ scales with the system's dimension, spanning the lattice. The dashed lines show the corresponding fits to the collected means; error bars denote the standard errors in the estimated means arising from fluctuations in $\langle \mathcal{S}_i\rangle$.}  
\label{fig:FractalDim_NthLargest}
\end{figure*}

\subsection{A Deterministic Percolation Transition In The Logistic GOL}\label{sec:cluster_analysis}

Motivated by the asymptotic behavior of the size of the largest cluster in the logistic GOL, we here complement these findings by studying the cluster dynamics of the system as $\lambda$ approaches $\lP=0.86055$ from below. By investigating the sizes and geometrical properties of clusters, we find that $\lP$ is the critical point of a percolation transition that separates two distinct phases of asymptotic behavior: a dynamic phase with no spanning cluster ($\lambda\leq\lP$) and a dynamic phase with a giant vacuum cluster that spans the lattice ($\lP<\lambda<\lA$).

We examine the size and geometrical properties of the largest clusters in the parameter range $\lambda \in [0.850, 0.875]$, where, for convenience, we focus on the five largest clusters (\MainFigref{fig:FractalDim_NthLargest}). First, we note that the highest-ranked clusters, i.e. clusters with $\{\langle \cS_1\rangle,...,\langle \cS_5\rangle\}$, are all composed of zero states (see top and bottom panels in \MainFigref{fig:FractalDim_NthLargest}a). When $\lambda\simeq 0.85$, all clusters exhibit comparable sizes but remain small relative to the lattice size (\MainFigref{fig:FractalDim_NthLargest}b). As $\lambda$ increases and approaches $\lP$, the size of each cluster increases, and the size of the second largest cluster reaches maximum at $\lambda=\lP$ (purple curve in \MainFigref{fig:FractalDim_NthLargest}b). When $\lambda > \lP$, the size of the largest cluster increases as they merge (red regions in \MainFigref{fig:FractalDim_NthLargest}a and inset in \MainFigref{fig:FractalDim_NthLargest}b), while the sizes of the lower-ranked clusters drop significantly.  

\subsubsection{Critical evolution of cluster capacity dimension}

Next, we analyze how the shapes of the largest clusters evolve with $\lambda$ through their capacity dimensions.   This `probes' whether clusters become scale-invariant near $\lP$.  Employing the box-counting method, the capacity dimension $\dc$ of the clusters is given by:
\begin{equation}
    \dc = - \lim_{\epsilon \to 0^+} \frac{\log \mathcal{N}(\epsilon)}{\log \epsilon}
\end{equation}
where $\mathcal{N}(\epsilon)$ denotes the minimum number of boxes of size $\epsilon$ needed to cover the cluster (see \SupNoteref{sec:box counting} for details). A capacity dimension of $\dc\sim1$ indicates that cluster shapes are more chain-like and sparse, while $\dc\sim2$ indicates that clusters are more area-like and dense. The capacity dimensions of the largest five clusters are plotted against $\lambda$ in \MainFigref{fig:FractalDim_NthLargest}c. The obtained $\dc$ converge towards each other as $\lambda$ approaches $\lP$ from below, signifying scale invariance at criticality. But once $\lambda > \lP$, the capacity dimensions diverge strongly: the largest cluster's capacity dimension increases, while the other clusters' capacity dimensions decrease.

Moreover, we examine how different samplings of the same clusters change at each \(\lambda\) by calculating the standard deviation \(\sigma_c\) of the capacity dimension. This allows us to quantify the stability of the shapes within a given cluster distribution. As shown in \MainFigref{fig:FractalDim_NthLargest}d the \(\sigma_c\) of every cluster decreases and reaches its minimum as $\lambda$ approaches \(\lP\) from below. When $\lambda > \lP$, the largest cluster's standard deviation remains stable over different samplings while \(\sigma_c\) of the other clusters increase. Such fluctuations are also reflected in the susceptibility profile seen in \MainFigref{fig:actcluster}b.

From the above analysis, it is evident that the highest-ranked clusters undergo a percolation transition at $\lP$. Indeed, as $\lambda$ approaches $\lP$ from below, the capacity dimensions of all clusters increase: they attain the same value ($\dc\approx 1.610$) and a minimal standard deviation ($\sigma_c \approx 0.01$). In other words, when $\lambda=\lP$, the shape of a cluster at a given time point is similar to the shape of any other cluster at any time point. Therefore, the clusters tend towards the same shape with respect to each other and only differ in size, providing strong evidence for scale invariance at $\lP$. Then, when $\lambda > \lP$, the shapes of the clusters change drastically with respect to each other (\MainFigref{fig:FractalDim_NthLargest}c) and with respect to their different samples (\MainFigref{fig:FractalDim_NthLargest}d). Here, the increase of the largest cluster's size and capacity dimension (dark red in \MainFigref{fig:FractalDim_NthLargest}a-c) signals the percolation transition, while the sizes and capacity dimensions of the smaller clusters decrease as they become smaller and more chain-like.

\subsubsection{Critical scaling of cluster sizes}

To further support the hypothesis that the transition at $\lP$ is percolation-like, we investigate how the largest cluster's size $\expect{\cS_1}$ scales with the lattice size $N$ as we approach $\lP$ (\MainFigref{fig:FractalDim_NthLargest}e), where we find that the scaling is the same as in classical percolation models~\cite{Stauffer1994,sahimi1994applications}. While relegating the details to \SupNoteref{sec:lambda_1_d_f_analysis}, here we report the observed scaling relationships:
\begin{equation}
    \expect{\cS_1(N)} \sim
    \begin{cases}
        \log N, & \text{for } \lambda < \lP, \\
        N^{\df}, & \text{for } \lambda = \lP,  \\
        N^D, & \text{for } \lambda > \lP,
    \end{cases}
\end{equation}
where $D=2$ denotes the dimension of the system, and $\df<D$ defines what is referred to as the fractal dimension. In the `subcritical' regime ($\lambda < \lP$), the largest cluster grows logarithmically with system size (left panel in \MainFigref{fig:FractalDim_NthLargest}e), meaning that there can be no giant cluster spanning the lattice. At the critical point ($\lambda = \lP$), the largest cluster follows a fractal scaling, reflecting the self-similar nature of the percolating cluster (\MainFigref{fig:FractalDim_NthLargest}e middle panel). Moreover, the same self-similarity induces pronounced fluctuations in $\langle \mathcal{S}_i\rangle$—visible in the error bars—, as clusters are drawn from a heavy-tailed distribution that follows a power law (see \MainFigref{fig:lambda_1_KS}).

The fitted fractal dimension is \( \df \approx 1.628 \) with a standard deviation of \( \sigma_f \approx 0.122 \), attributed to deviations from the exact critical point \( \lP \) and standard error of $\expect{\mathcal{S}_i }$.  Moreover, the identified \( \df \) and \( \dc \) values are mutually consistent, both falling within the same uncertainty range. In the `supercritical' regime ($\lambda > \lP$), the largest cluster grows with the system dimension ($D=2$), indicating the formation of a percolating cluster that spans the lattice (\MainFigref{fig:FractalDim_NthLargest}e right panel). 

Besides the scaling relations governing $\cS_1(N)$, the scalings of lower-ranked clusters $\expect{\cS_i(N)}$ indicate that, as $N\rightarrow\infty$, $\expect{\cS_i(N)}$ diverge to infinity only at the critical point $\lambda=\lP$ (\MainFigref{fig:FractalDim_NthLargest}e) (see \SupNoteref{fig:SizeScaling} for details). Taken together, all the analyses of largest clusters (\MainFigref{fig:FractalDim_NthLargest}) indicate the emergence of a percolating cluster and a phase transition~\cite{newman2018networks} at $\lambda=\lP$.

\subsubsection{Wrapping probabilities for precise determination of the percolation threshold}\label{sec:wrappingprob}

In the context of percolation, the wrapping probability is a key dimensionless observable, and we now use it to find the thermodynamic percolation point $\lP(N\to\infty)$. As grid size $N$ increases one expects to see more percolation events for $p > p_c$, due to the higher probability of sampling the dominant state forming the percolation cluster. Similarly, for $p < p_c$ the percolation events appear less frequently for larger $N$, since there is no dominant cluster in this region and the probability that a randomly chosen cluster is percolating becomes smaller and smaller.
For dimensionless percolation properties this leads to the finite-size scaling form \cite{Stauffer1994}:
\begin{equation}
    \Pi(p) = \Phi\!\big[(p - p_c) f(N)\big],
\end{equation}
which predicts a fixed point for the observable, since
$\Pi(p_c) = \Phi(0)$ is independent of $N$. Since we have the same expectation for the probability of observing
percolation events above and below the threshold in our model, we
expect the same fixed-point behavior, with the replacement
$p \to \lambda$ and $p_c \to \lP(N\to\infty)$.

In numerical simulations this dimensionless property $\Pi$ is best
captured by the wrapping probability $R_\mathrm{W}$ and is used to locate the
percolation point of the infinite lattice. To achieve this we consider four different wrapping-probability forms \cite{Newman_2001}:

\begin{figure}[ht!]
\centering
\includegraphics[width=\linewidth]{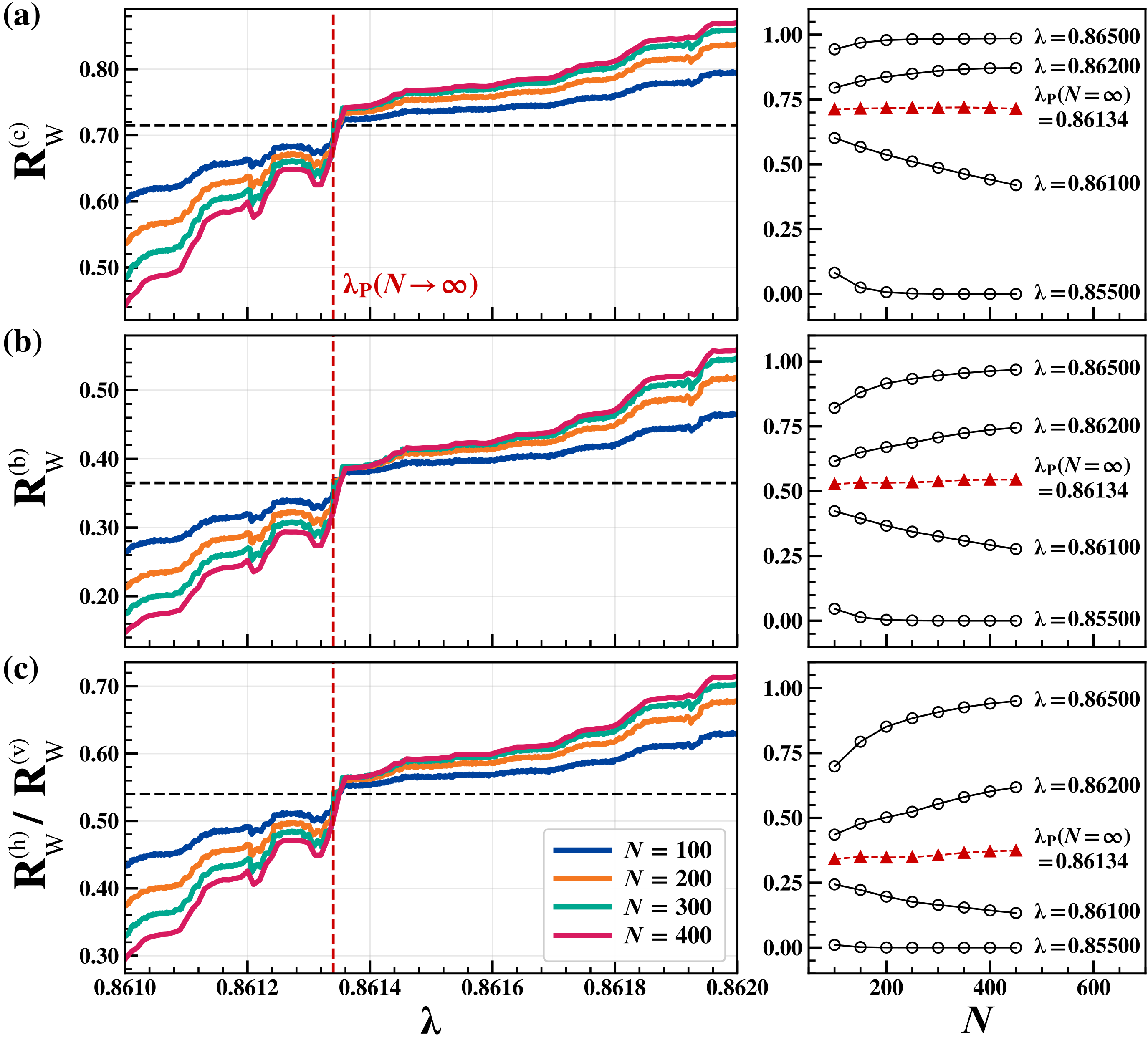}
\caption{\textbf{Wrapping probabilities around the percolation transition.}
\textbf{(a)} Wrapping probability in either the horizontal or vertical
direction, $\RW{e}$.
\textbf{(b)} Wrapping probability in both directions simultaneously,
$\RW{b}$.
\textbf{(c)} Wrapping probabilities in the horizontal and vertical
directions, $\RW{h}$ and $\RW{v}$.
The first column shows the evolution of the corresponding
wrapping probabilities as a function of $\lambda$, while the second column shows their evolution as a function of system size $N$ for
selected $\lambda$ values chosen around the thermodynamic percolation point $\lP(N \to \infty)$. The curves for different system sizes $N$ intersect at a common fixed point
$\lambda_P(N \to \infty) \simeq 0.86134$ (red dashed line), with corresponding
wrapping probabilities marked at $\RW{e} \approx 0.7163, \RW{b} \approx 0.3667,
\RW{h} = \RW{v} \approx 0.5415$ (black dashed lines).}

\label{fig:wrappingProb}
\end{figure} 
\begin{itemize}
    \item $\RW{e}$ is the probability that there exists a cluster
    which wraps in at least one direction, i.e.\ either horizontally, vertically, or both.

    \item $\RW{b}$ is the probability that there exists a cluster
    which wraps simultaneously in both the horizontal and vertical
    directions.

    \item $\RW{h}$ and $\RW{v}$ are the probabilities that there exists a cluster which wraps around the system in the horizontal  and vertical directions, respectively. Due to symmetry, these  probabilities are equal in our system.

\end{itemize}

Plotting these quantities in \MainFigref{fig:wrappingProb}(a-c), we indeed
observe the expected finite-size scaling with a common fixed point at $\lP(N \to \infty)$.
The curves for values of $\lambda$ in the vicinity of this fixed point are found to obey the following scaling relations:
\begin{equation}\label{wrapping_finitsizescaling}
\begin{cases}
\dfrac{\partial R_\mathrm{W}(\lambda,N)}{\partial N} > 0, & \text{for } \lambda > \lP,\\[4pt]
\dfrac{\partial R_\mathrm{W}(\lambda,N)}{\partial N} \approx 0, & \text{for } \lambda = \lP,\\[4pt]
\dfrac{\partial R_\mathrm{W}(\lambda,N)}{\partial N} < 0, & \text{for } \lambda < \lP.
\end{cases}
\end{equation}
From the common crossing point of the curves at different $N$, we
identify the thermodynamic percolation threshold as
$\lP(N \to \infty) \simeq 0.86134$. Accordingly, the finite-size
shift of the $\mathcal{S}_i$ peak is expected to move the peak position towards $\lP \to 0.86134$ as $N \to \infty$ (see \SupNoteref{sec:Second_largest_divergence} for details).

Since $R_\mathrm{W}(\lP(N\to \infty))$
is also independent of system size, it defines a dimensionless
universal quantity that depends only on the system. It is therefore natural to record the observed values of
$\RW{h}, \RW{v}, \RW{e}$, and $\RW{b}$, which are:
\begin{equation}
\begin{aligned}
\RW{e} &\approx 0.7163,\\
\RW{b} &\approx 0.3667,\\
\RW{h} &= \RW{v} \approx 0.5415.
\end{aligned}
\end{equation}
which satisfy the relation \cite{Newman_2001}
$\RW{e} = \RW{h} + \RW{v} - \RW{b}$, so that only two of
the wrapping probabilities are independent.

\FloatBarrier
\subsection{Cluster Size Distributions Near The Critical Points}\label{sec:cluster_size}
 
Having previously established the scaling properties of the largest clusters with system size, we next investigate the extent to which cluster size distributions near the critical points \(\lP\) and \(\lA\) follow power laws. To do this, we perform numerical simulations of the logistic GOL to compute the distribution of cluster sizes, $p(\cS)$, in the vicinity of each critical point.

As a brief overview, at $\lP=0.86055$, $p(\cS)$ seems to follow a power law, while for other nearby $\lambda$ values, distributions appear as stretched exponentials (see \SupNoteFigref{fig:log_for_different_lambda}a). On the other hand, there are multiple $\lambda$ values close to $\lA=0.875$ where the distributions are reminiscent of power laws, but only if the largest vacuum clusters are disregarded (see \SupNoteFigref{fig:log_for_different_lambda}b). While relegating technical aspects of the computation of $p(\cS)$ to \SupNoteref{sec:cluster size distributions}, below we leverage quantitative methods to test whether such distributions are indeed best described by power laws.

The scaling of power-law data is rarely valid across the entire domain of cluster sizes. More often, the power law applies only for values greater than some lower bound $\cS_\mrm{min}$, i.e., only the `tail' follows a power law. In such cases, the cluster size distribution is expected to follow:
\begin{equation}\label{eq:dicrete_poweR_Waw}
    p(\cS) = \frac{\cS^{-\tau}}{\zeta(\tau, \cS_\mrm{min})} \quad \text{for} \quad \cS \geq \cS_\mrm{min}
\end{equation}
where $\tau$ is the power-law exponent (the Fisher exponent \cite{fisher1969}), $\cS_\mrm{min}$ is the lower cutoff, and $\zeta(\tau, \cS_\mrm{min})$ denotes the generalized zeta function
\begin{equation}
    \zeta(\tau, \cS_\mrm{min}) = \sum_{\cS=\cS_\mrm{min}}^{\infty} \cS^{-\tau} = \sum_{\cS=0}^{\infty} (\cS+\cS_\mrm{min})^{-\tau}.
\end{equation}
The corresponding complementary cumulative distribution function (cCDF) then reads:
\begin{equation}
    \mathcal{F}(\cS) = \sum_{\cS'=\cS}^{\infty} p(\cS') = \frac{\zeta(\tau, \cS)}{\zeta(\tau, \cS_\mrm{min})}.
\end{equation}
Using the numerically computed cCDF, we determine the Fisher exponent $\tau$ and the lower bound $\cS_\mrm{min}$ by employing the Kolmogorov-Smirnov (KS) method \cite{kolmogorov1933,smirnov1948table,Clauset2009, alstott_powerlaw_2014, barabasi2013network}. In addition to finding the optimal values of $\tau$ and $\cS_\mrm{min}$, the KS method assesses how well the power-law model fits to the data in comparison to other fat-tailed distributions (see \SupNoteref{sec:KS method} for details).

\begin{figure}[t!]
\includegraphics[width=\columnwidth]{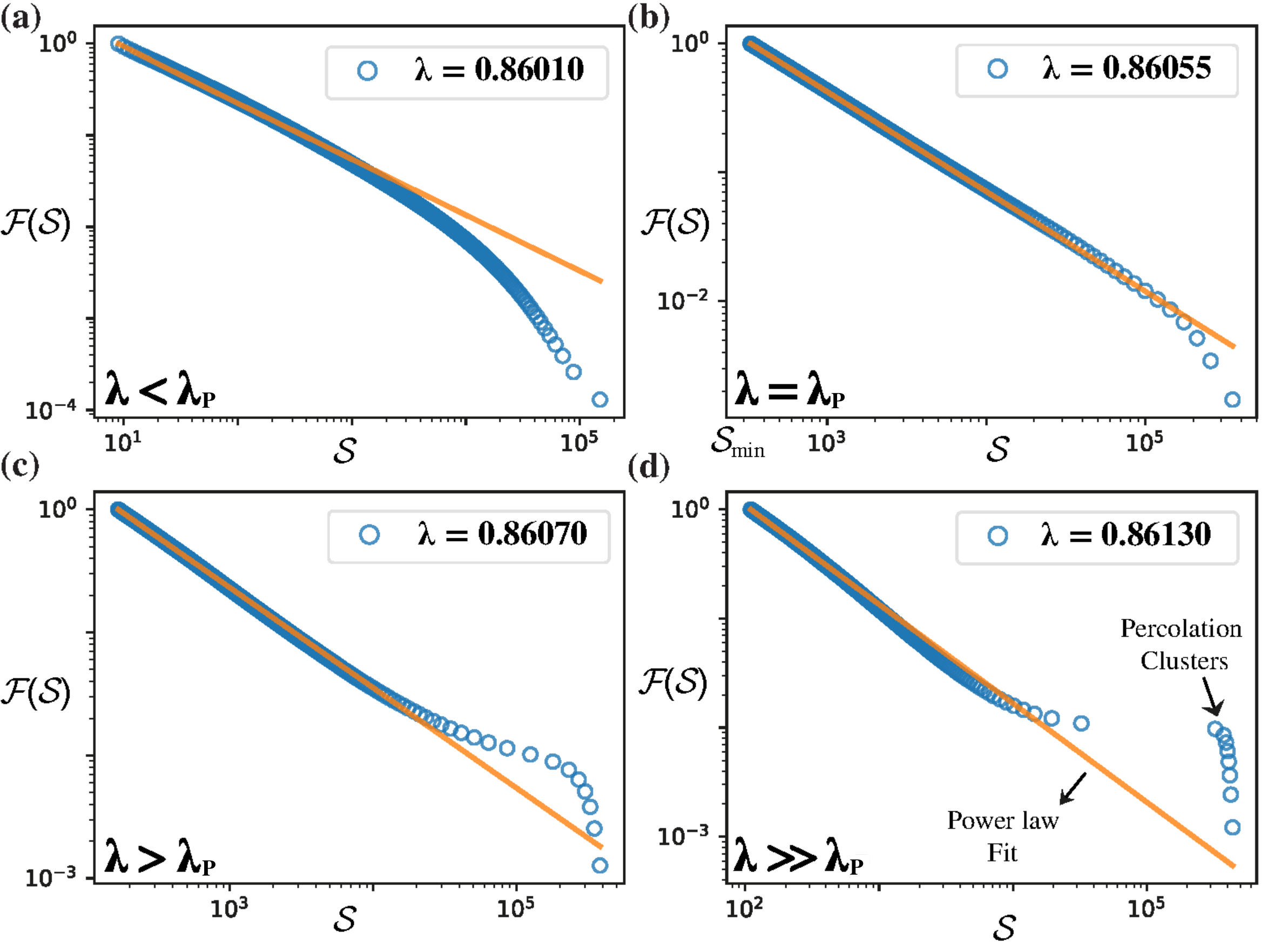}
\caption{\textbf{Behavior of cluster size distribution around $\lP = 0.86055$}. The empirical complementary cCDFs with logarithmic-binning are shown in blue, with the fitted power-law in orange, for $\lambda$ values \textbf{(a)} below, \textbf{(b)} close, and \textbf{(c-d)} above $\lP$. The x-axis starts from the optimal $\cS_\mrm{min}$ determined by the KS method. The cluster size distribution becomes a power law (with exponential cutoff) only very close to the critical point $\lP$.}
\label{fig:lambda_1_KS}
\end{figure}

Initially, we apply the KS method to the cluster size distributions for $\lambda$ values in the vicinity of $\lP$. 
In this range, the logistic GOL unlocks dynamic control over cluster behavior, enabling precise tuning of the tail fatness in the cluster distribution through the tuning parameter $\lambda$. 
As shown in \MainFigref{fig:lambda_1_KS}a, the tail of the cCDFs undershoots the power-law line when $\lambda < \lP$. As $\lambda$ increases, the number of zero clusters and the variance of cluster sizes increase, resulting in a fatter tail (\MainFigref{fig:lambda_1_KS}b). 
However, at $\lambda = \lP$, we observe that the tail fits with a power law with exponential cutoff (\MainFigref{fig:lambda_1_KS}b). 
This cutoff is due to finite-size effects (see \SupNoteref{sec:SizeEvolution} for details). Further increases in $\lambda$ lead to the loss of perfect linearity of the cluster size distribution (\MainFigref{fig:lambda_1_KS}c), supporting the assertion that $\lP$ is the critical point for the emergence of a giant cluster. 
Beyond this point, the largest vacuum cluster separates from the rest of the distribution and begins to percolate. 
As \(\cS_1\) grows to be comparable to the system size \(N^2\), it diverges from the main body of the distribution (arrow in \MainFigref{fig:lambda_1_KS}d). 
This evolution is also evident through simulation snapshots in \MainFigref{fig:FractalDim_NthLargest}a. 

The fit results of the KS method for cluster distribution at $\lambda=\lP$ yield a power-law distribution with exponential cutoff, with the following coefficients:
\begin{align}
    \lP &= 0.86055: \begin{cases}
        \tau = 1.81 \pm 0.03 \\
        \cS_\mrm{min} = 560 \pm 150
    \end{cases}  
    \label{lambda_1_fit_results}
\end{align}
The plausibility of the optimal power-law fit to the numerical data is supported by the Kolmogorov--Smirnov (KS) test, and the relation \( \tau = 1 + \df / 2 \) \cite{Hu2016} also holds within the uncertainty regime. Additionally, the log-likelihood ratio test determines whether alternative fat-tailed distributions (e.g., exponential, stretched exponential, or log-normal) offer a better fit than the power law. The power-law distribution with an exponential cutoff best characterizes the system at $\lP$ (as detailed in \SupNoteTableref{table:statconclusion}).

Next, we discuss the cluster size distributions in the vicinity of $\lA$. As previously mentioned, in this range the lattice is dominated by the largest percolating cluster. However, we find that the distribution of other smaller clusters exhibits interesting behavior. Therefore, when applying the KS method to the cluster size distributions near $\lA$, we always neglect the largest cluster by `trimming out' the separated part of the distribution (arrow in \MainFigref{fig:lambda_1_KS}).

\begin{figure}[ht!]
\includegraphics[width=\columnwidth]{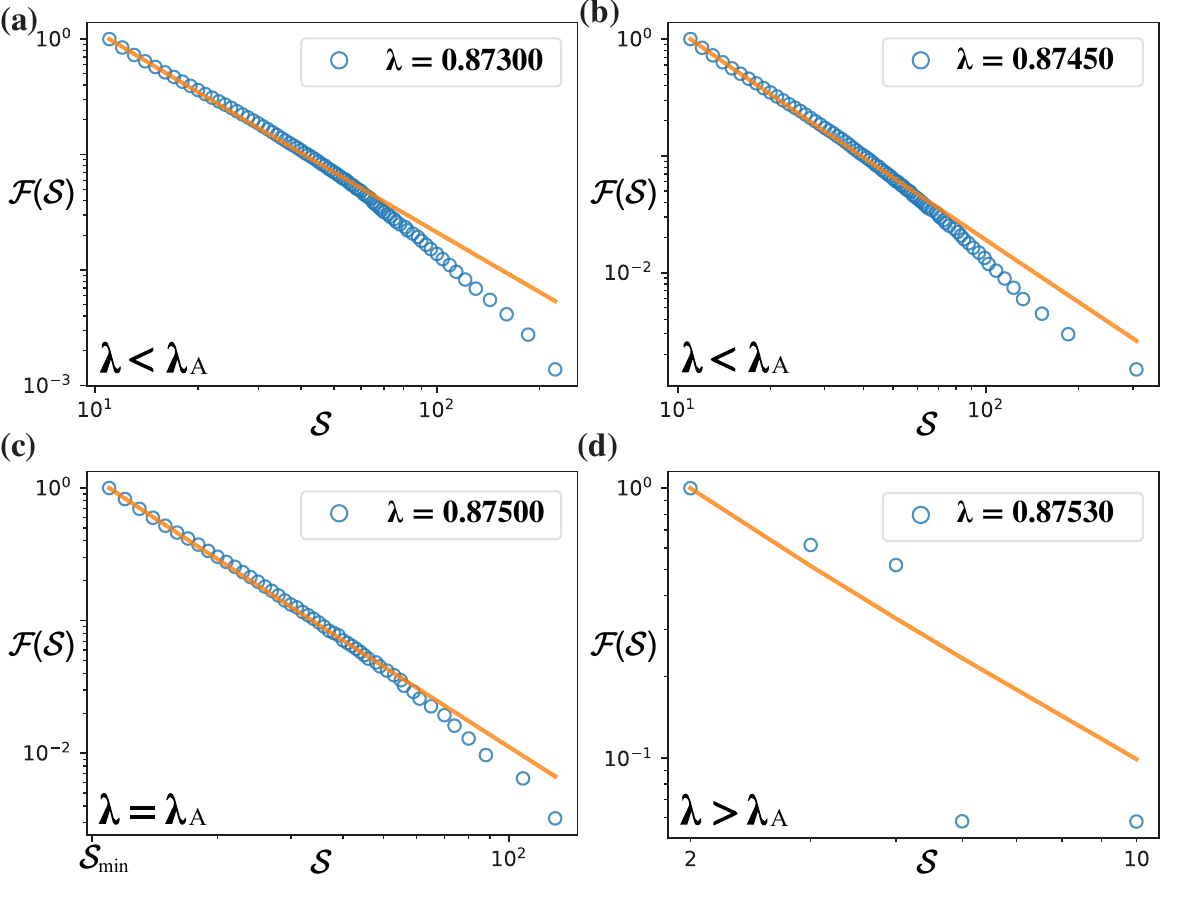}
\caption{\textbf{The evolution of cluster size distribution around $\lA = 0.875$, when the largest vacuum cluster is disregarded.} The cCDFs with logarithmic-binning are shown in blue, with the fitted power law in orange. The largest vacuum clusters are ignored here and the x-axis starts from the optimal $\cS_\mrm{min}$ determined by KS method. \textbf{(a-c)} The distribution evolves near $\lP$ by approaching a power law from below and becomes a power law near $\lA$. For $\lambda > \lA$ \textbf{(d)}, asymptotic activity and cluster formation cease, leading to the disappearance of cluster dynamics.}
\label{fig:lambda_2_KS}
\end{figure}

As $\lambda$ approaches $\lA$ from below, the trimmed cluster size distribution displays similar behavior to that in the vicinity of $\lP$ (\MainFigref{fig:lambda_2_KS}). In \MainFigref{fig:lambda_2_KS}a, the distribution is best described by a power law with cutoff, while in \MainFigref{fig:lambda_2_KS}b--c it follows a power law without cutoff. As $\lambda$ increases above $\lA$, the system enters an inactive phase, resulting in the disappearance of cluster dynamics (\MainFigref{fig:lambda_2_KS}d). The corresponding power-law parameters at $\lA$, obtained from the KS method, are:
\begin{align}
    \lA &= 0.875: \begin{cases}
        \tau = 2.9 \pm 0.1  \\
        \cS_\mrm{min} = 11 \pm 3 \quad .
    \end{cases}   
\end{align}
See \SupNoteTableref{table:statconclusion} for model comparison test details.

Further statistical analyses using the KS method over different parameter values in the range $0.8 < \lambda < 0.9$ are discussed in \SupNoteref{sec:goodness_of_fit}, where we evaluate the quality of the power-law fits for the cluster size distributions near the critical points. The results of these statistical analyses indicate that:
\begin{itemize}
    \item At $\lP = 0.86055$, the tail of cluster size distribution follows a power law with exponential cutoff.
    \item At $\lA = 0.875$, the tail of cluster size distribution follows a pure power law when the giant vacuum cluster is disregarded.
\end{itemize} 

The different natures of criticality at $\lP$ and $\lA$ are also reflected in their Fisher exponents, $\tau$. At $\lambda=\lP$, where $1 < \tau < 2$, the mean and all higher moments diverge, including the mean $\expect{\cS(\lP)}$. This arises because the percolation behavior causes the bulk of the distribution to be highly concentrated in the tail. In the thermodynamic limit, the tail of the distribution (\MainFigref{fig:lambda_1_KS}c) extends to infinity, resulting in $\expect{\cS(\lP)} \to \infty$. In contrast, at $\lambda=\lA$, where $2 < \tau < 3$, the mean remains finite, and only the variance and higher moments diverge. This means that, unlike $\lP$, the critical behavior at $\lA$ does not consist of clusters comparable to the system size. Below, we discuss the potential mechanisms involved in the emergence of such power-law distributions.

\subsubsection{Contrasting mechanisms for criticality from cluster size distributions}

The mechanism behind the percolation transition can be explained as follows. As $\lambda$ approaches $\lP$ from below, the system promotes more zero states because several neighborhood sums increase from $m\leq t_3$ to $m>t_3$, inducing decay instead of growth. As a consequence, clusters of quiescent states grow continuously with $\lambda$ until they merge with each other at $\lambda=\lP$. In this respect, the dominance of zero states in the grid and the power law behavior of cluster sizes indicate that $\lP$  marks the point of a deterministic percolation transition. We moreover note that the cluster size distribution exponent $\tau \simeq 1.81$ found at $\lP$ is lower than exponents in classical 2D ordinary percolation models ($\tau>2$), hence the universality class of this transition remains unclear. However, we also note that extremely similar exponents have been observed in interesting scenarios, such as the no-enclave percolation model~\cite{Sheinman2015} and 2D random walk~\cite{Federbush2021}. This exponent was previously interpreted as characterizing holes within the cluster backbone on the dual lattice~\cite{Hu2016}. In our model, we believe the observed clusters correspond to such backbone clusters, as the decay process---governed by the Moore neighborhood---naturally eliminates vacuum-enclosed active regions. Specifically, in the thermodynamic limit, any site surrounded entirely by 0-states is forced to decay, effectively producing a ``no-enclave'' behavior.

We propose that such unconventional exponents have not been observed in classical percolation models because the no-enclaving mechanism introduces a form of radial anisotropy that traditional site or bond percolation models cannot capture. The Moore neighborhood provides a minimal and illustrative realization of this radial anisotropy: the central site is influenced by its eight surrounding neighbors, yet it does not exert influence back in a strictly symmetric manner. This inherent asymmetry introduces an inward directional bias that is absent in conventional percolation frameworks. Our model thus serves as a paradigmatic example demonstrating how unconventional Fisher exponents with $\tau < 2$ can naturally emerge in physical systems through the incorporation of radial anisotropy.   

\begin{figure}[t!]
\includegraphics[width=\columnwidth]{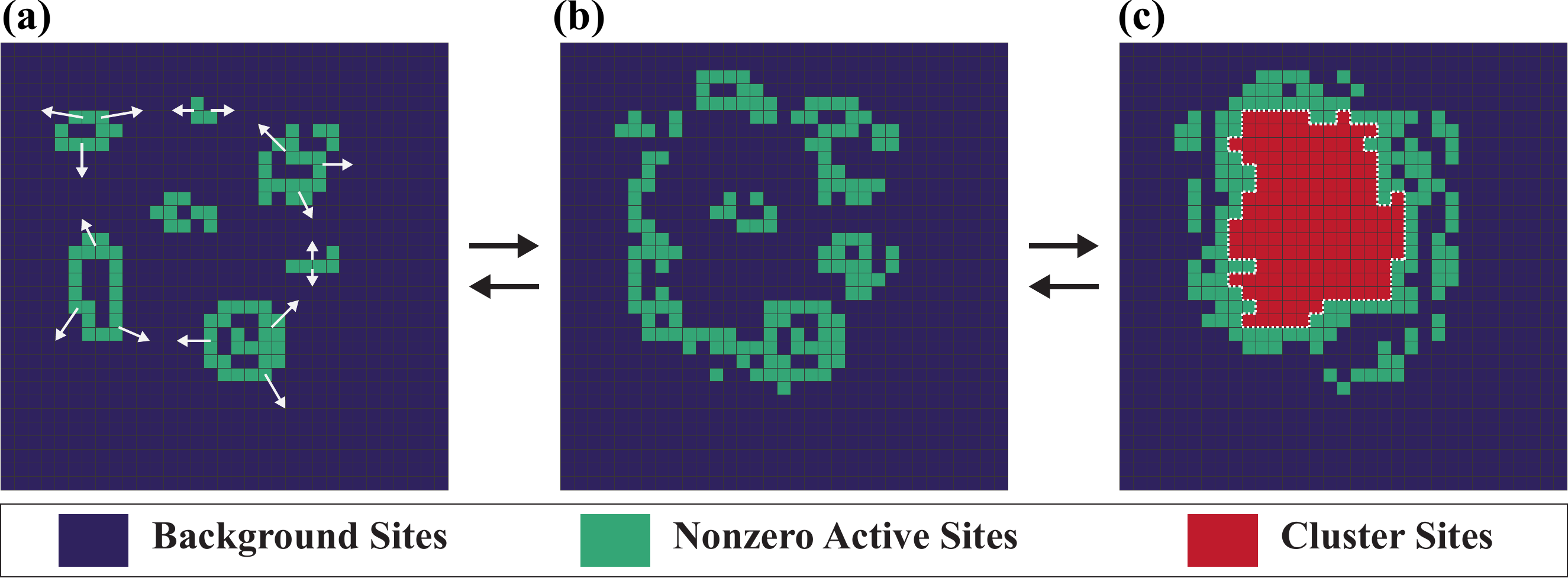}
\caption{\textbf{A peculiar form of deterministic, self-organized criticality in the vicinity of $\lA$}. As $\lambda$ approaches $\lA$ from below, the \textbf{(a)} active nonzero cells `move' in a manner that \textbf{(b)} encircles the zero-state background cells and \textbf{(c)} occasionally forms clusters with the associated power law behavior. These steps are two-way: just as clusters are formed, they can also fragment in the same manner.}
\label{fig:SOC}
\end{figure}

The mechanism for the emergence of power law around $\lA$ is fundamentally different from $\lP$. Around $\lA$, the system is dominated by a vacuum cluster of quiescent states that serves as a `playground' for activity with diverging susceptibility. This susceptible activity spreads in a particular fashion such that it `encircles' quiescent regions in the grid, giving rise to smaller zero-state clusters (\MainFigref{fig:SOC}). Interestingly, the size of these zero-state clusters encircled by activity follows a power-law distribution, which emerges close to $\lA$. Beyond this value, the power-law ceases because the asymptotic activity stops due to more neighborhood sums inducing decay instead of growth.

\subsubsection{Possible self-organized criticality in the vicinity of the $\lA$}\label{subsec:selforganized}

We believe that the power-law behavior in the vicinity of $\lA$ reflects a form of self-organized criticality (SOC), similar to the one discussed by \citet{bak1989self}. In these studies, it has been shown that the activity clusters follow a power law when the asymptotic state of Conway's GOL is continually perturbed by altering single sites. In our case, the power-law distribution of zero clusters occurs at multiple points in the region $\lambda \rightarrow \lA^-$ (see \MainFigref{fig:lambda_2_KS}a-b and \SupNoteref{sec:goodness_of_fit}), suggesting scale invariance over a continuous parameter range, similar to the SOC behavior. We speculate that, around \( \lA \), the logistic GOL administers `perturbations' to itself continually via neighborhood configurations of $m = 4\lambda$ and $m = 5-4\lambda$ (\MainTableref{tab:transitions-neighborhood-main}). Such configurations seem to be occurring frequently enough to maintain a persistent activity in the lattice through cascades of nearby state changes, thereby generating activity profiles that propagate throughout the lattice. Such activity shares similar nonlinearity with Conway's GOL, but in contrast, it is persistent without the need for any external perturbations. In this context, if Per Bak's system operates in a `stimulated' SOC regime, our system functions in a `spontaneous' SOC regime.

\section{Discussion}\label{sec:discussion}

Scale-invariant dynamics is a striking phenomenon emerging in a large variety of spatially extended systems. Such complex systems, despite being defined by local interactions, happen to display units of equal states that organize in clusters with no characteristic size and/or duration. While scale invariance appears either in the form of self-organized criticality, or in the form of parameter-driven criticality, systems displaying such behavior are typically associated with random external inputs (e.g., random `grains' of sand added in the sandpile model~\cite{bak1987self}), or with probability in interactions (e.g., temperature in the Ising model~\cite{Parisi2002,Creswick_1997,Talapov_1996}), suggesting that stochasticity is an essential ingredient for such critical behavior. Here, we revisit this idea by investigating a purely deterministic update rule and control parameter that nevertheless display scale-invariant dynamics in both of its typical forms, and show that deterministic criticality can also emerge in a manner similar to classical systems involving stochasticity. Unlike other models that pair deterministic rules with probabilistic control parameters, our system has deterministic update rules and a deterministic control parameter. Initial conditions appear as a source of randomness; however (as detailed in \SupNoteref{sec:invariancetoinitialconds}), for a broad range of initial densities the system relaxes to the same $\lambda$‑dependent stationary statistics, so the random initial condition acts only as a transient start‑up and does not constitute a control parameter.

Specifically, we identify critical behavior in the asymptotic dynamics of the logistic GOL, an extension of Conway's GOL where a single parameter ($\lambda$) tunes the rate at which sites change in every iteration~\cite{Ibrahimi2019}. Using numerical simulations of the system, we identify three distinct dynamical regimes separated by two critical deterministic phase transitions. In the first one (phase I), where $\lA<\lambda\leq1$, the asymptotic dynamics of the logistic GOL is virtually the same as Conway's GOL, with long transients that eventually settle to sparse populations of stable/oscillating structures in a spanning vacuum cluster of quiescent states~\cite{Bagnoli1991}. The second dynamical regime (phase II) lies between $\lP<\lambda\leq\lA$, where the logistic GOL becomes asymptotically active -- i.e. the dynamics persists in the thermodynamic limit -- but still with a vacuum cluster that spans the lattice. As $\lambda$ decreases further, activity increases and the size of the vacuum cluster is consequently reduced. The size of this cluster decreases with $\lambda$ until it disconnects into smaller clusters at $\lP$. This second transition defines the limit of the third dynamical regime (phase III), $\lambda\leq \lP$ where the logistic GOL is active and there is no vacuum cluster spanning the lattice.

We use standard measures from percolation theory to study the dynamics of largest clusters close to the critical point ($\lP$) separating phases II and III, and find clear numerical evidence for a deterministic percolation transition hidden in the Game of Life. We believe that there are two aspects that make this transition interesting. First, the study of percolation transitions -- which are widespread in models of physics, networks, and population dynamics -- is particularly uncommon in systems where clusters are generated by purely deterministic interactions. We are only aware of a few spatially extended~\cite{CHATE1988,GRASSBERGER1991,CUCHE1997} systems where transitions from homogeneous to chaotic behavior have been compared to directed percolation processes. Second, the cluster size distribution at $\lambda=\lP$ has a Fisher exponent of $\tau\simeq 1.81 < 2$, which is also not typical for percolating systems \cite{Sheinman2015,Federbush2021}. While previous works explain $\tau<2$ in terms of backbone clusters \cite{Hu2016}, our results show that radial anisotropy in purely local updates can reproduce the same effect.

Moreover, we study the system in the vicinity of the transition between phases I and II, and find that $\lA$ marks the transition point between these phases. We find that this transition is defined by a discontinuity in the asymptotic activity, and is not related to any cluster merging process. However, we observe that the activity profiles near the border of the active asymptotic phase, i.e. when $\lambda \rightarrow \lA^-$, give rise to clusters of zero-states that follow a power-law distribution (\MainFigref{fig:SOC}). Our findings suggest that this behavior reflects a peculiar form of self-organized criticality, related to the one observed in early studies of Conway's GOL~\cite{bak1989self,Alstrom1994}. Yet, the self-organized criticality observed in $\lambda \rightarrow \lA^-$ is spontaneous, i.e., it does not require external inputs in order to showcase scale-invariant clusters. In this respect, it would be interesting to find other models exhibiting the same kind of behavior and identify general underlying mechanisms of such criticality. 

Overall, our study highlights the idea that scale-invariant dynamics is not limited to complex systems with stochasticity in their interactions. Specifically, we provide evidence that percolation transitions occurring in deterministic systems are similar to their counterparts observed in other classical complex systems. In contrast, our transition exhibits an unconventional exponent that departs from standard hyperscaling expectations typically valid for equilibrium systems. Moreover, we show that there are systems that exhibit self-organized criticality spontaneously in their dynamic, asymptotic attractor states, and that do not require external perturbations to display this kind of behavior. Our results are consistent with a form of self‑organized criticality that lies beyond traditional stochastic and Abelian frameworks, and suggest the possibility of a broader class of SOC‑like behavior. Taken together, while Conway’s GOL does not refer to any particular real-world system, we believe that the scale-invariant dynamics revealed here is inherently related to the physics of critical phenomena and will incite new studies on deterministic physical models.

\FloatBarrier 

\section{Data Availability}

All data generated or analyzed in this study are available from the corresponding authors upon reasonable request. Selected portions are publicly accessible at the project repository: \url{https://github.com/HakanAkgn/ClusterAnalyzer/tree/main/Paper_Data}.

\section{Code Availability}

We provide a general-purpose, open-source library for cluster analysis and criticality detection to support the broader research community, available at \url{https://github.com/HakanAkgn/ClusterAnalyzer}. The library enables analysis and visualization of cluster distributions and dynamics, implementation of power-law testing, extraction of key critical exponents (e.g., fractal dimension and Fisher exponent), and a range of related analyses.

\FloatBarrier
  
\section{Acknowledgments}

H.A.\ acknowledges the support from A\*STAR (SIPGA), T{\"U}B{\.I}TAK~2205, and the NUS Research Scholarship. 
T.T.\ acknowledges the computing services provided by TRUBA. CHL acknowledges support by the Ministry of Education, Singapore (Tier-II award number: MOE-T2EP50222-0003 and Tier-I WBS code: A-8002656-00-00). S.J.\ acknowledges the computing services provided by UHEM.

\section{Author contributions}
H.A. conceived the study, performed all theoretical/numerical calculations, and drafted the manuscript. X.Y. performed cluster-size distribution analyses, optimized the codebase, managed the repository, and contributed to writing and revision. T.T. contributed to visualization, writing and revision. M.I. contributed to conceptualization, writing and revision. S.J. initiated the research direction and supervised the project and contributed to writing and revision. C.H.L. supervised the project and contributed to writing and revision. All authors discussed the results and commented on the manuscript.
  
\clearpage 
\twocolumngrid   
\beginsupplement

\section{Model Implementation and Simulation Details}\label{sec:LGOL_implementation}
\begin{figure}[ht!]
\centering
\includegraphics[width=1\linewidth]{camera_figures/Figure_9.pdf}
\caption{\textbf{State space across different orders and truncation process.} \textbf{(a)} The state space in the logistic GOL expands into a Cantor set, where the dissection ratio is set by $\lambda$. Different levels show the emergence of the first two orders of the Cantor set from applying the rules to a uniform distribution. In this study, we truncate the state space at a finite order to analyze the cluster dynamics of the system. \textbf{(b)} The decay and growth operations for the first order are illustrated in the truncated space. The truncation preserves the operational regimes, and lumping nearby states allows for maintaining the same dynamic behavior as the logistic GOL. }
\label{fig:statespace}
\end{figure}

\begin{figure}[ht!]
\centering
\includegraphics[width=\linewidth]{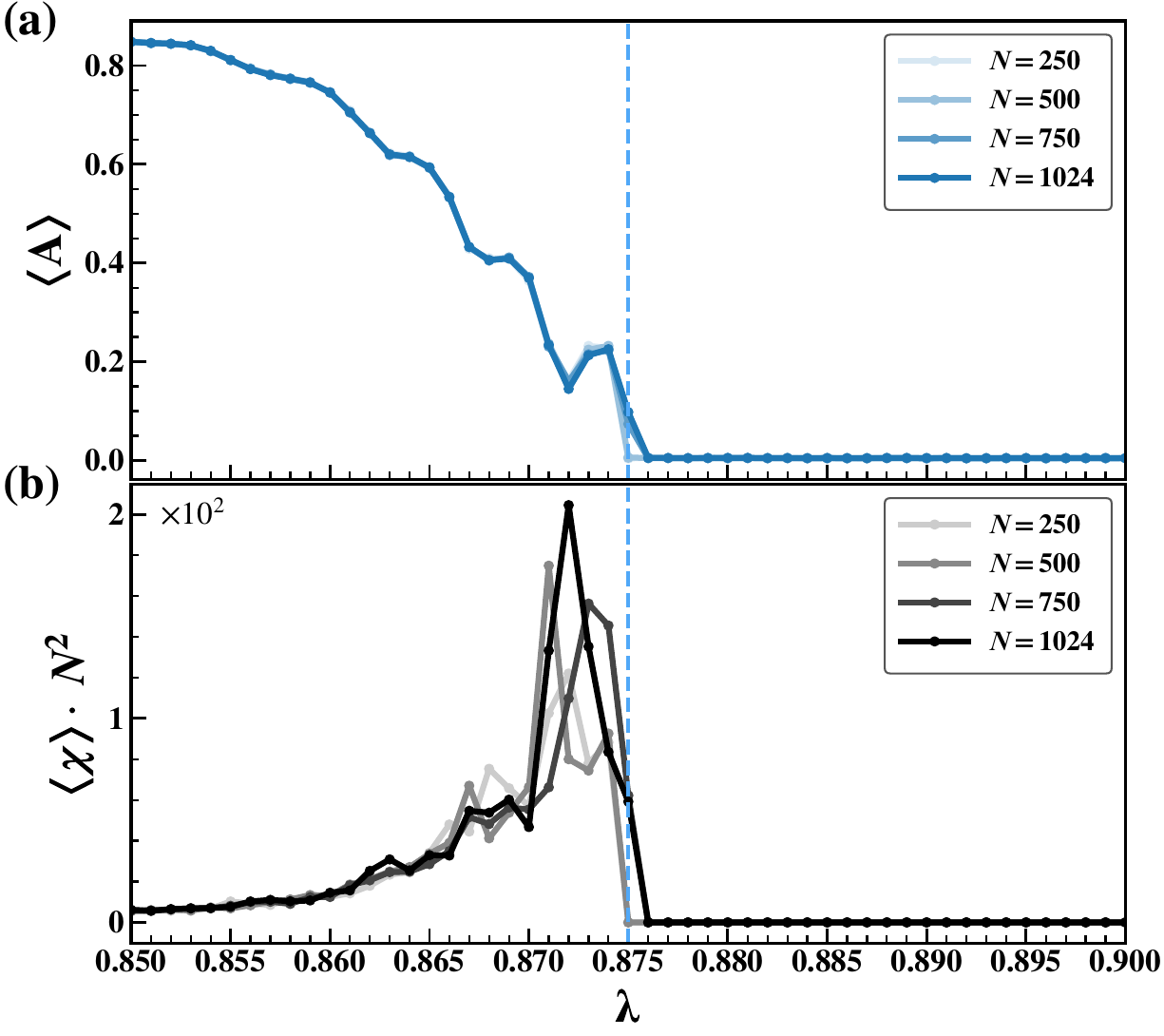}
\caption{\textbf{System-size analysis of $\lA$.}
The dashed vertical line marks the common threshold $\lA$.
\textbf{(a)} Asymptotic mean activity $\langle A\rangle$ versus $\lambda$ for system sizes $N$, and the curves collapse, indicating negligible size dependence and consistent inactivity of the system for $\lambda > \lA$. \textbf{(b)} Susceptibility $\langle \chi\rangle$ exhibits similar behavior across $N$ with a common peak location; minor shape differences reflect statistical uncertainty (larger standard errors) in $\langle \chi\rangle$ near the peak.}

\label{fig:activity_chi_vs_lambda_bysize}
\end{figure}

Starting from the logistic GOL, we retain only the Cantor set up to order \(n\), truncating the rest. The state space is discretized into  $|L_n| = 2^{n+1}$ states, which are generated recursively as follows:

\[
L = 
\begin{cases}
L_0 = \{0, 1\}  \\
L_1 = \{0, 1-\lambda, \lambda, 1\}  \\
L_2 = \{0, (1-\lambda)^2 \dots 1 - (1-\lambda)^2,  1\}    \\
\vdots  \\
L_n = (1 - \lambda)L_{n-1} \cup \left( L_{n-1}(1-\lambda)+\lambda \right) 
\end{cases}
\]

The set $L_n$ represents all possible combinations of growth and decay operations on the initial set $L_0 = \{0, 1\}$. To ensure the preservation of the number of states, we map the state space onto itself:
 
\begin{gather}
s^{t+1}_j =
\begin{cases}
    \mathbf{S} s^{t}_j \equiv s^{t}_j & \text{if } t_1 \leq m^{t}_j < t_2, \\
    \mathbf{G} s^{t}_j \equiv M_{L_n} \left((1 - \lambda)s^{t}_j + \lambda\right) & \text{if } t_2 \leq m^{t}_j < t_3, \\
    \mathbf{D} s^{t}_j \equiv M_{L_n} \left((1 - \lambda)s^{t}_j\right) & \text{otherwise.}
\end{cases}
\end{gather}

where \( M_{L_n} \) is defined as the nearest-element projection onto \( L_n \):

\[
M_{L_n}(x) = \operatorname*{arg\,min}_{y \in L_n} |x - y|
\]

This setup with \( M_{L_n} \) ensures that each transformed state is mapped to the nearest valid state within \( L_n \), preserving the structure and permutation of the state space. The evolution and truncation of state spaces across different orders are illustrated in \SupNoteFigref{fig:statespace}a.

The operational domains remain the same throughout the truncation process, but now the range of Moore sum $m$ is discretized instead of being continuous. This modification makes it possible to perform cluster analyses of the system while maintaining the features of logistic GOL with an un-truncated Cantor set. The growth/decay operations for the first-order truncation are illustrated in \SupNoteFigref{fig:statespace}b. As the order $n$ tends to infinity, the truncated version approaches the continuous state space of the logistic GOL. We note here that, when performing numerical simulations, the state space is nevertheless truncated in some order depending on the numerical resolution of the implementation program.  

The numerical parameters used in all simulations of the truncated logistic GOL are summarized in Table~\ref{tab:sim_params}. Unless stated otherwise, we use these parameters throughout: a $1024\times1024$ lattice with periodic boundary conditions, a burn-in of $10^{5}$ time steps followed by averaging over a further $10^{5}$ time steps, and multiple independent realizations for each value of the control parameter $\lambda$. For plots at a single system size $N$, we perform $500$ simulations per value of $\lambda$, whereas for plots comparing multiple system sizes we perform $100$ simulations per value of $\lambda$ for each $N$. Clusters and their sizes are obtained by connecting adjacent cells that are in the same state, realized by the union-find algorithm \cite{Tarjan1975}. All the expected values of observables (activity, susceptibility, cluster sizes, etc.) are acquired by time-averaging and ensemble-averaging the raw data. Additionally, different initialization densities were tested and found to converge to the same thermodynamic behavior, provided the initial density supports a persistent activity.

\begin{table}[t]
\centering
\caption{\textbf{Simulation parameters.} Unless stated otherwise, the values listed here are used throughout the paper.}
\label{tab:sim_params}
\begin{tabular}{ll}
\hline\hline
Quantity & Value / Description \\
\hline
Lattice size & $1024 \times 1024$ \\
Boundary conditions & Periodic (PBC) \\
Burn-in time & $T_{\mathrm{burn}} = 10^{5}$ time steps \\
Averaging window & $T_{\mathrm{avg}} = 10^{5}$ time steps \\
Simulations per $\lambda$ (single $N$ plots) & $n_{\mathrm{sims}} = 500$ \\
Simulations per $\lambda$ (multi $N$ plots) & $n_{\mathrm{sims}} = 100$ \\
\hline\hline
\end{tabular}
\end{table}

\section{System-Size Analysis of $\lA$}\label{sec:SizeEffectsOnA}

Across $\lambda \in [0.85, 0.88]$ (step size $0.001$), observables associated with $\lA$ exhibit negligible dependence on system size, as shown in \SupNoteFigref{fig:activity_chi_vs_lambda_bysize}. 
In \SupNoteFigref{fig:activity_chi_vs_lambda_bysize}a, the asymptotic mean activity $\langle A\rangle$ for different $N$ nearly collapses onto a single curve and indicate that the system is consistently inactive for $\lambda > \lA$. 
In \SupNoteFigref{fig:activity_chi_vs_lambda_bysize}b, the peak structure of $\langle \chi\rangle$ is likewise consistent across sizes; residual shape differences are attributable to larger standard errors near the peak (see \SupNoteref{sec:standarderror}).

\begin{table}[t]
\centering
\caption{\textbf{Explicit operational transitions for $\lP$.} The table above presents the fifth-order $\lambda$ neighborhood transitions for the $t_1$ and $t_3$ neighborhoods at $\lP$. The panel below shows neighborhoods undergoing transition, while the lower left panel illustrates the numerical evolution of these neighborhoods as $\lambda$ varies between $0<\lambda<1$.}
\label{tab:explicit_lambda_P_neighbohoodTable}

\begingroup 
\setlength{\tabcolsep}{2pt}%

\begin{tabular}{|c|p{0.77\columnwidth}|}
\hline
\textbf{Transition} & \textbf{Neighborhood} \\ \hline

$\mathrm{S}\leftrightarrow\mathrm{D}$ &
$\displaystyle
t_{1}=
\begin{array}[t]{@{}l@{}}
2\times[0] + [-(\lambda-1)^5] \\
+\,2\times\big[(\lambda-1)\big(\lambda\big((1-\lambda)^3-1\big)+(\lambda-1)^3\big)\big] \\
+\big[-(\lambda-1)(\lambda^2-\lambda+1)\big] + [1-\lambda] \\
+\big[\lambda^4-3\lambda^3+3\lambda^2-\lambda+1\big]
\end{array}
$ \\ \hline

$\mathrm{G}\leftrightarrow\mathrm{D}$ &
$\displaystyle
t_{3}=
\begin{array}[t]{@{}l@{}}
[-(\lambda-1)^5] \\
+\big[(\lambda-1)\big(\lambda\big((1-\lambda)^3-1\big)+(\lambda-1)^3\big)\big] \\
+\,2\times\big[-(\lambda-1)(\lambda^2-\lambda+1)\big] + [1-\lambda] \\
+\,2\times\big[\lambda^4-3\lambda^3+3\lambda^2-\lambda+1\big] \\
+\big[-\lambda^5+4\lambda^4-6\lambda^3+4\lambda^2-\lambda+1\big]
\end{array}
$ \\ \hline

\multicolumn{2}{|c|}{%
\begin{minipage}{\columnwidth}
\vspace{1mm}
\centering
\includegraphics[width=\linewidth]{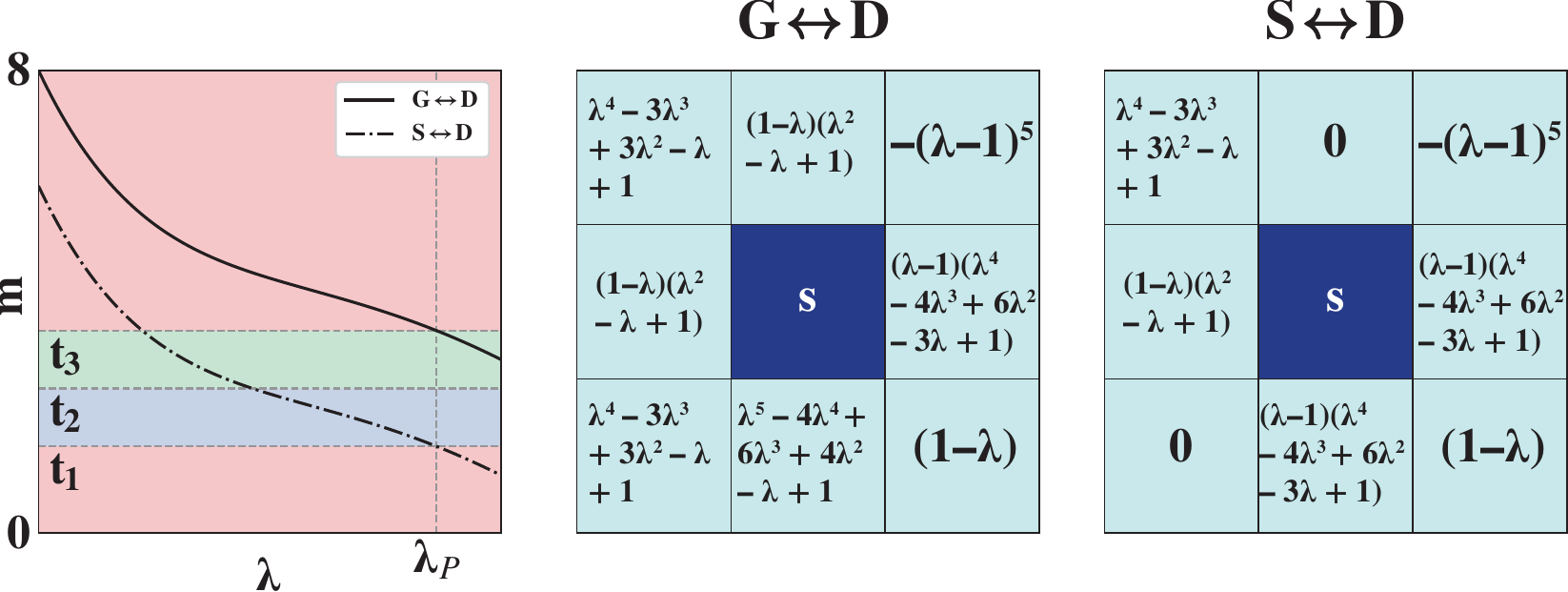}
\vspace{0.2mm}
\end{minipage}
} \\ \hline

\end{tabular}

\endgroup
\end{table}

\begin{figure}[ht!]
\centering
\includegraphics[width=1\linewidth]{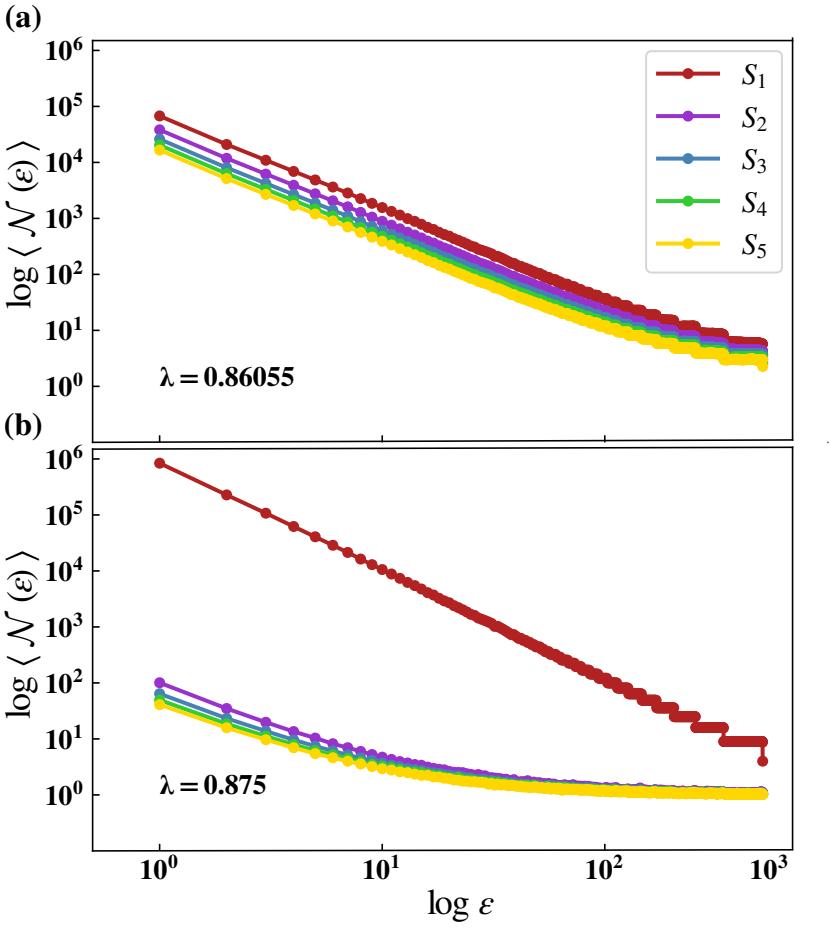}
\caption{\textbf{The capacity dimension obtained by box-counting}. The plots show the (averaged) box counts \(\langle N(\epsilon) \rangle\) v.s. box sizes \(\epsilon\), \textbf{(a)} for \(\lambda = 0.86055\) and \textbf{(b)} for \(\lambda = 0.875\). The capacity dimension equals the negative of the slope near \(\epsilon = 0\). As \(\lambda\) increases from \(\lambda = 0.86055\) to \(\lambda = 0.875\), the largest cluster gradually separates from the rest, exhibiting an increasing capacity dimension, while the other clusters' capacity dimensions decrease.}

\label{fig:BoxCounting}
\end{figure}

\begin{figure}[ht]
\centering
\includegraphics[width=1\linewidth]{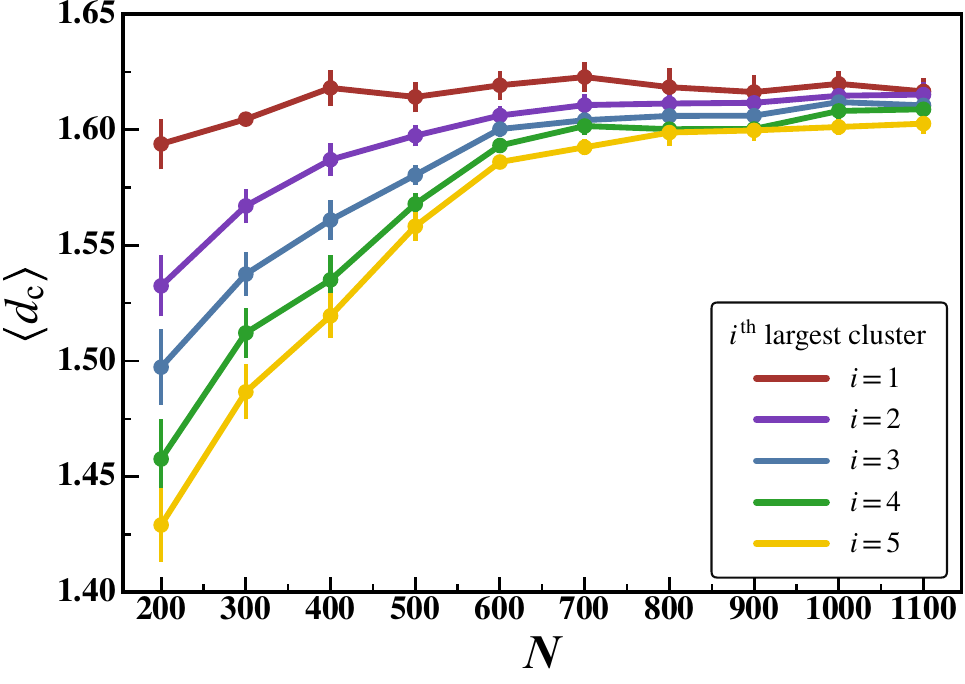}
\caption{\textbf{Evolution of capacity dimension with system size.} For the largest cluster, the capacity dimension $\dc$ is stable at large grid sizes. As system size increases, $\dc$ for lower-ranked clusters likewise stabilizes, exhibits smaller standard errors, and converges to a value minimally different from that of the largest cluster.}
\label{fig:capacityDimbysize}
\end{figure}

\section{Explicit Neighborhood of Percolation Transition $\lP$}\label{sec:explicit_lambda_P_neighborhood}

To approximate a target value, the algorithm selects a subset of Cantor set states that sum closely to the desired target within a specified tolerance. This is achieved using a branch-and-bound approach \cite{LandDoig1960}, which explores possible combinations of states while discarding unpromising paths. In this approach, the algorithm iteratively builds subsets of the Cantor set by adding states and checking if the current sum is within tolerance. The process is optimized by pruning paths that cannot meet the target, based on the following criteria:

\begin{itemize}
    \item \textbf{Subset Size Constraint}: Paths that exceed the allowed number of states are discarded.
    \item \textbf{Tolerance Check}: Paths with cumulative sums that deviate from the target by more than the tolerance are also discarded. The tolerance is set to 0.00001 to match the resolution of our numerical simulations.
    \item \textbf{Feasibility Pruning}: The algorithm estimates the minimum and maximum possible sums with remaining states. Paths are pruned if they cannot reach or exceed the target based on these bounds.
\end{itemize}

This process ensures efficient exploration of feasible subsets, yielding an optimal selection that best approximates the target. Accordingly, \SupNoteTableref{tab:explicit_lambda_P_neighbohoodTable} represents the 5th-order Cantor set. As the percolation transition is continuous, approaching the exact percolation point $\lP$ with high decimal precision requires progressively higher-order neighborhoods.  \SupNoteTableref{tab:explicit_lambda_P_neighbohoodTable} presents the summed and simplified polynomial representations of these neighborhoods. As similar higher-order polynomials change their operational regions, neighborhood characteristics change, and cluster dynamics progressively evolve. It should be noted that this selected state evolution over $\lambda$ serves as an illustrative example of how changes in the operational region influence state dynamics and, eventually, cluster evolution. It does not represent an exact transition, as $\lP$ shifts towards $\lP(N\to\infty)$ as $N$ increases and higher-order neighborhoods can always be found within the Cantor set.

\section{Numerical Methods for Cluster Characterizations}

\subsection{Box-counting method for the capacity dimension}\label{sec:box counting}

The box-counting method determines the capacity dimension of an object by covering it with grids of varying box sizes and counting the number of boxes, \(N(\epsilon)\), that contain part of the object. By analyzing how \(N(\epsilon)\) changes with the box size \(\epsilon\), the capacity dimension \(\dc\) is obtained through the following steps:

\begin{enumerate}
    \item Cover the cluster with a grid of boxes of size \(\epsilon\).
    \item Count the minimal number of boxes needed to cover the cluster, denoted as \(\mathcal{N}(\epsilon)\).
    \item Repeat the steps above over multiple time steps and different initializations to obtain the average box count \(\expect{\mathcal{N}(\epsilon)}\) for each box size ($\epsilon)$).
    
    \item Plot \( \log \expect{\mathcal{N}(\epsilon)}\) v.s. \(\log \epsilon\).
    \item Determine the slope of the plot in the small box size region (specifically  $\epsilon= [1,10]$). The capacity dimension \(\dc\) is given by:
    \[
    \expect{\mathcal{N}(\epsilon)} \propto \epsilon^{-\dc} \quad \rightarrow \dc = -\lim_{\epsilon \to 0} \frac{\log \expect{\mathcal{N}(\epsilon)}}{\log \epsilon}
    \]
\end{enumerate}

The slopes representing the capacity dimensions of the five largest clusters at two critical points are shown in \SupNoteFigref{fig:BoxCounting}, highlighting their distinct characteristics. Near $\lP$ all clusters have the same slope \SupNoteFigref{fig:BoxCounting}a, showing increased self-similarity of the system. Above $\lP$, the largest cluster's slope increases and becomes more area-like, while other cluster slopes decrease and become more chain-like \SupNoteFigref{fig:BoxCounting}b. As the largest cluster percolates and fills the entire PBC grid (excluding quiescent states), it forms a two-dimensional surface with $\dc = 2$.  

However, it should be noted that the box-counting behavior holds only until the box size reaches the size of the clusters. Similar to other percolation models \cite{christensen2005complexity}, this relationship can be understood in terms of the mass of a given cluster at the percolation threshold: 

\begin{equation}
M(C_i, \lP; \ell) = S_i(\lP; \ell) \propto
\begin{cases}
\ell^{\dc} & \text{for } \ell \ll R_s, \\
S_i & \text{for } \ell \gg R_s,
\end{cases}
\end{equation}

where \( S_i(\lP; \ell) \) is the number of sites in the \( i \)-th cluster for a given window length \( \ell \), which corresponds to the effective box size in the counting process. When \( \ell \) exceeds the characteristic cluster radius \( R_s \), further increasing the window size (i.e., the effective box size) does not capture additional cluster sites; instead, the larger boxes simply encompass the existing sites, leading to no increase in the count of occupied boxes. This is because the cluster is now fully covered, meaning that regardless of additional window size increases, the same number of boxes is needed to cover the entire cluster. This results in a flattening behavior, as seen in \SupNoteFigref{fig:BoxCounting}b, where smaller clusters are fully covered by a constant number of boxes.

Moreover, to assess the stability of the capacity dimension, we plot $\langle \dc\rangle$ computed at $\lP$ versus system size in \SupNoteFigref{fig:capacityDimbysize} (error bars denote the standard error). As the grid size increases, the estimated $\dc$ remains stable, and the $\dc$ values of lower-ranked clusters converge to values that differ only minimally from that of the largest cluster, as expected. The deviations from this common value at small grid sizes arise primarily from a finite-size shift of the percolation point toward lower $\lambda$ (see Sec.~\ref{sec:Second_largest_divergence}).

\subsection{The scaling fits and fractal dimension at $\lambda=\lP$}\label{sec:lambda_1_d_f_analysis}

\SupNoteFigref{fig:FractalDim_NthLargest}(e) presents the numerical fits for the scaling of cluster sizes \( \langle \cS_i(N) \rangle \) across different percolation regimes. We perform fits on data points for system sizes \( N \) from 200 to 1300 in increments of 50. A moderate system size, such as \( N = 200 \), ensures statistically consistent cluster dynamics across various \( \lambda \) neighborhoods, independent of initial configuration. Since \( \lambda = 0.86055 \) does not exactly match the analytical percolation point and has additional significant decimal places beyond \( 0.00001 \), it is expected that, like other percolation models, system scaling will eventually deviate from a perfect power law \cite{Stauffer1994}. 

The reported standard deviation combines two contributions: the standard error of the mean (the error bars in \MainFigref{fig:FractalDim_NthLargest}e) and the change in the fitted exponent when $\lambda$ is varied within $\lP \pm 0.000005$. Since common neighborhoods tend to decay and become passive in this regime, smaller grid sizes are sufficient for effectively capturing scaling dynamics. For these fits, we specifically use the $\lambda$ values: $\lambda = 0.855$, $\lambda = 0.86055$, and $\lambda = 0.865$, respectively, for the subcritical, critical, and supercritical regimes. In the subcritical regime (\MainFigref{fig:FractalDim_NthLargest}a), the cluster sizes follow a logarithmic scaling law $\expect{\cS_i(N)} \sim a_i \log N + b_i$, with coefficients $a_i$ and $b_i$ depending on the cluster rank $i$.    

At the critical point $\lP$, the scaling transitions to a power law, with the largest cluster following $\expect{\cS_1(N)} \sim N^{1.628}$, indicative of the system's fractal nature at criticality. Subleading clusters scale similarly with different exponents. These power-law fits reflect the fractal dimensions of clusters ($\df$), a hallmark of critical phenomena. While the largest cluster follows a fractal scaling law, subleading clusters exhibit different exponents as a result of their sensitivity to system size and non-system-spanning nature \cite{christensen2005complexity}. These clusters remain fractal, scaling with exponents indicative of their distribution near criticality. Numerical fits estimate the system's fractal dimension as \( \df \approx 1.628\) with a standard error of \( \sigf \approx 0.122 \), reflecting deviations from the critical point \( \lP \) in simulations. The most pronounced standard errors (error bars) appear in the critical regime, as expected, since at criticality the cluster-size distribution follows a heavy-tailed power law that amplifies variance relative to the subcritical and supercritical regimes.

In the supercritical regime, the largest cluster $\expect{\cS_1}$ scales with the system's Euclidean dimension, following a numerical fit of $\expect{\cS_1(N)} \sim N^{1.9997}$, which is very close to the expected $N^2$, signaling the emergence of a percolating cluster. Meanwhile, subleading clusters adhere to logarithmic-like scaling, indicating that while they grow with system size, they remain much smaller compared to the largest cluster. This analysis confirms that the scaling behavior of clusters across percolation regimes is consistent with classical percolation models.

\begin{figure}[t!]
\centering
\includegraphics[width=\columnwidth]{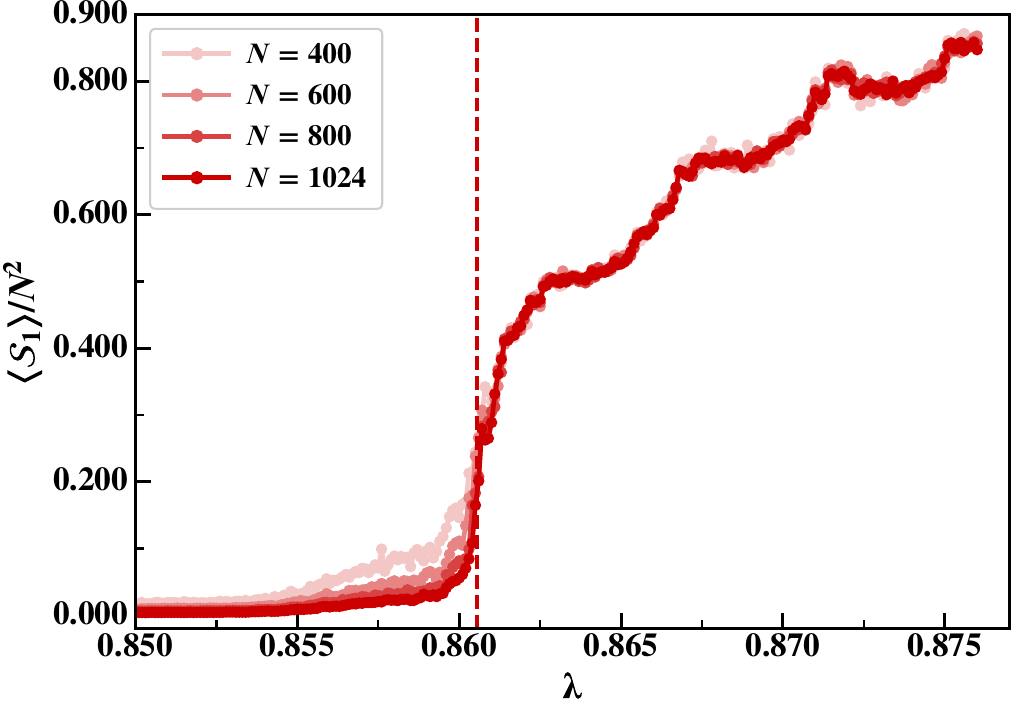}
\caption{\textbf{Largest-cluster size v.s. system size near $\lP$.} In the supercritical regime, the largest-cluster size $S_1$ is essentially stable. Finite-size effects are most pronounced at and below criticality, while the transition in $S_1$ sharpens around $\lP$ as system size increases.}
\label{fig:SizeScalingS1} 
\end{figure} 

\begin{figure*}[ht!]
\centering
\includegraphics[width=\textwidth]{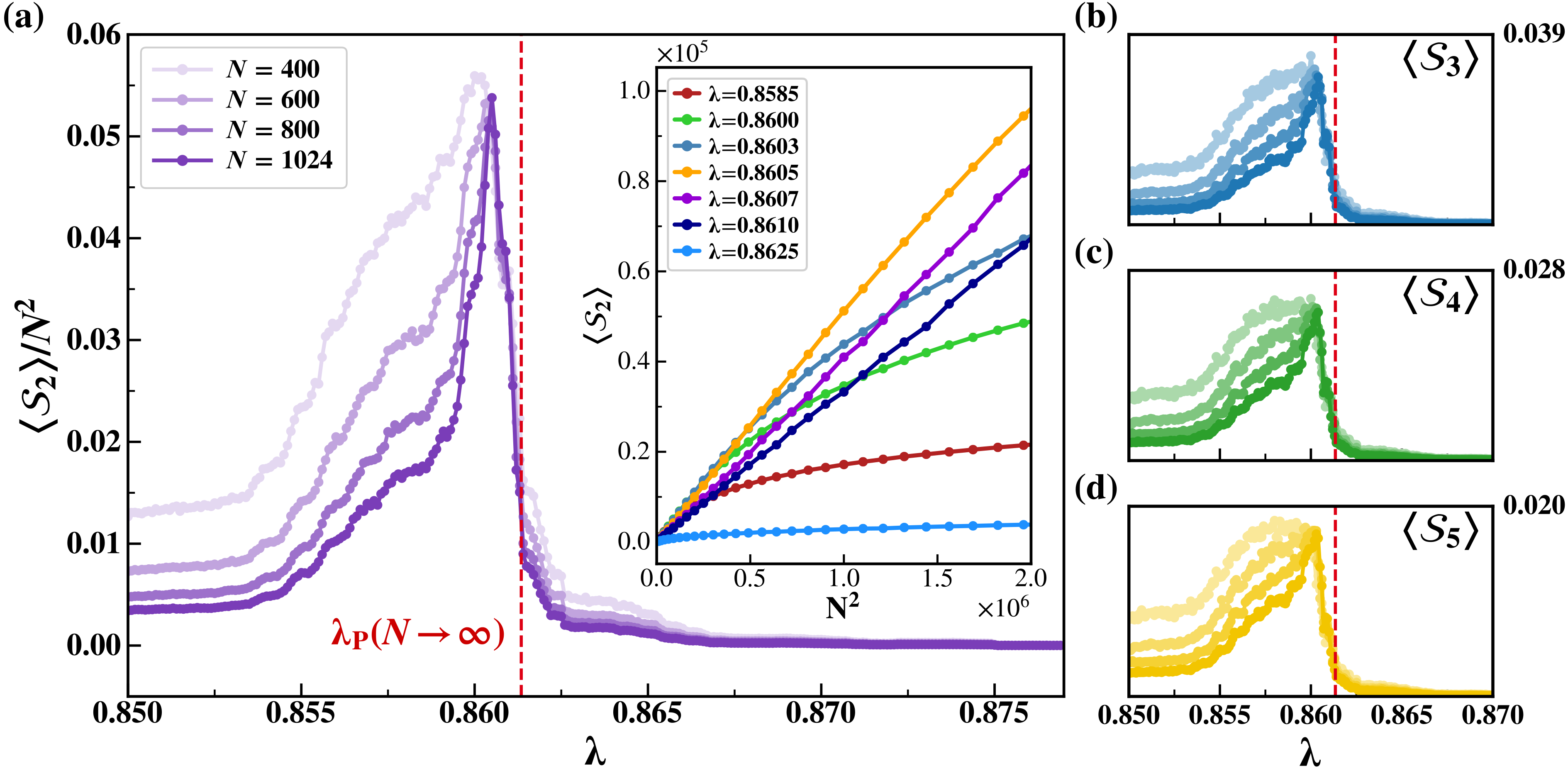}
\caption{\textbf{Size evolution of subleading clusters with respect to system size near $\lP$.} 
\textbf{(a)} As the system approaches $\lP$, the peak of $\langle \mathcal{S}_2 \rangle$ diverges with increasing $N$, appearing as an increasingly sharp peak that shifts rightward towards $\lP(N\to \infty)$. 
\textbf{(b–d)} Similar divergence and rightward-shifting peak sharpening are observed for $\langle \mathcal{S}_3 \rangle$, $\langle \mathcal{S}_4 \rangle$, and $\langle \mathcal{S}_5 \rangle$, reflecting the self-similar nature of the system.}

\label{fig:SizeScaling} 
\end{figure*}
\subsection{Evolution of cluster-size statistics with system size}\label{sec:Second_largest_divergence}

We examine how $\langle \cS_i(N) \rangle$ scales with system size $N$. The simulation was conducted over the range $\lambda \in [0.85, 0.88]$, with increments of 0.0001. It should be again noted that all top-ranking clusters are quiescent clusters.  

In \SupNoteFigref{fig:SizeScalingS1}, $S_1$ is largely insensitive to finite-size effects in the supercritical regime, while the subcritical side sharpens as the system approaches $\lP$, consistent with classical percolation models\cite{zhu2017finitesizescalingtheory}. This sharpening manifests as more pronounced peaks for lower-ranked clusters. As $\lP$ is approached, the second largest cluster $\langle \cS_2(N) \rangle$ grows rapidly, as shown in \SupNoteFigref{fig:SizeScaling}a. At $\lP$, it exhibits the fastest divergence, while at nearby points, the growth is slower. In the limit $N \rightarrow \infty$, $\langle \cS_2 \rangle$ shown in \MainFigref{fig:FractalDim_NthLargest}a diverges sharply at $\lP$, signaling the emergence of a percolating cluster and the phase transition~\cite{newman2018networks}. Likewise, the peaks of $\langle \mathcal{S}_i \rangle$ in \SupNoteFigref{fig:SizeScaling} (b--d) exhibit similar sharpening and divergence, reflecting the self-similar nature of the system.

Moreover, in line with percolation theory, the finite size of the lattice leads to a shift of the apparent percolation threshold
$\lP(N)$ away from its thermodynamic value. As the system size
increases, these shifts become progressively smaller and
$\lP(N)$ approaches the thermodynamic percolation threshold
$\lP(N \to \infty)$~\cite{sahimi1994applications,zhu2017finitesizescalingtheory},
which we identify with wrapping probabilities. Notably, although
$\lambda$ is a purely deterministic control parameter, its regulation
of the $0$-state configurations effectively mimics the role of the
probabilistic control parameter in stochastic percolation. This
correspondence naturally explains the observed rightward convergence
$\lP(N < \infty) \to \lP(N = \infty)$, directly analogous
to the probabilistic case where $p_c(N < \infty) \to p_c(N = \infty)$.

\subsection{Cluster size distributions from numerical simulations} \label{sec:cluster size distributions}
 
\begin{figure}[h!]
\centering
\includegraphics[width=1\columnwidth]{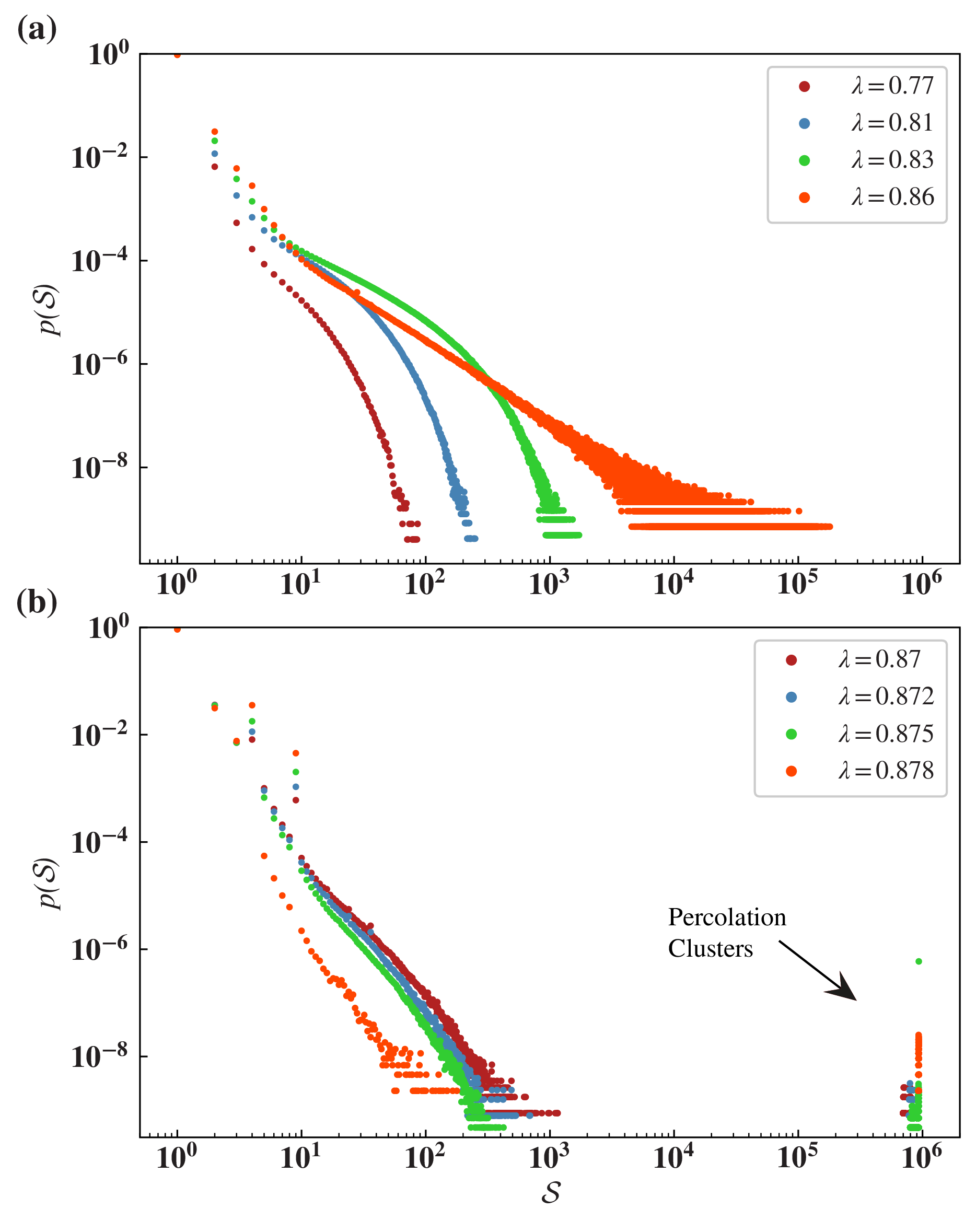}
\caption{\textbf{Log-log plot of the empirical probability density function (PDF) of cluster sizes for different $\lambda$ values.} \textbf{(a)} Evolution of PDFs towards $\lP$ from below shows that the PDF tails get fatter, both by shifting to higher cluster sizes and increasing the spread of the tail. At $\lP$, the tail extends up to the system size, regardless of grid size, indicating scale-invariance. Constrained by the system size, the fitted power-law has an additional exponential cutoff term ($x^{-\tau}\mathrm{e}^{-\lambda x}$). \textbf{(b)} Around $\lA$, samples of the percolation cluster separate from the rest, piling up away at the end shown with the black arrow. At $\lA$, the percolating cluster is discarded before fitting the power law, and the fitted model is a pure power law ($x^{-\tau}$). It doesn't have the cutoff term because only the percolating cluster is affected by the system size.}
\label{fig:log_for_different_lambda}
\end{figure}
We present the cluster size distributions near two critical points $\lP$ and $\lA$ in \SupNoteFigref{fig:log_for_different_lambda}. 
We obtain the numerical count of clusters using the union-find algorithm \cite{Tarjan1975}, and by normalizing these counts with the total number of clusters, we interpret the data as frequency distributions and treat them as probability density functions (PDFs), denoted by \( p(\cS) \). At $\lP=0.86055$, the cluster size distribution $p(\cS)$ seems to follow a power law, while others around it appear as stretched exponentials (\SupNoteFigref{fig:log_for_different_lambda}a). The piles at the far tail of the distributions around $\lA$ are contributed by the samples of percolating clusters denoted by the black arrow. After discarding the piles and truncating the lower curving head, the cluster size distribution at $\lA=0.875$ also appears to follow a power law (\SupNoteFigref{fig:log_for_different_lambda}b).

\begin{figure}[ht!]
\includegraphics[width=1\columnwidth]{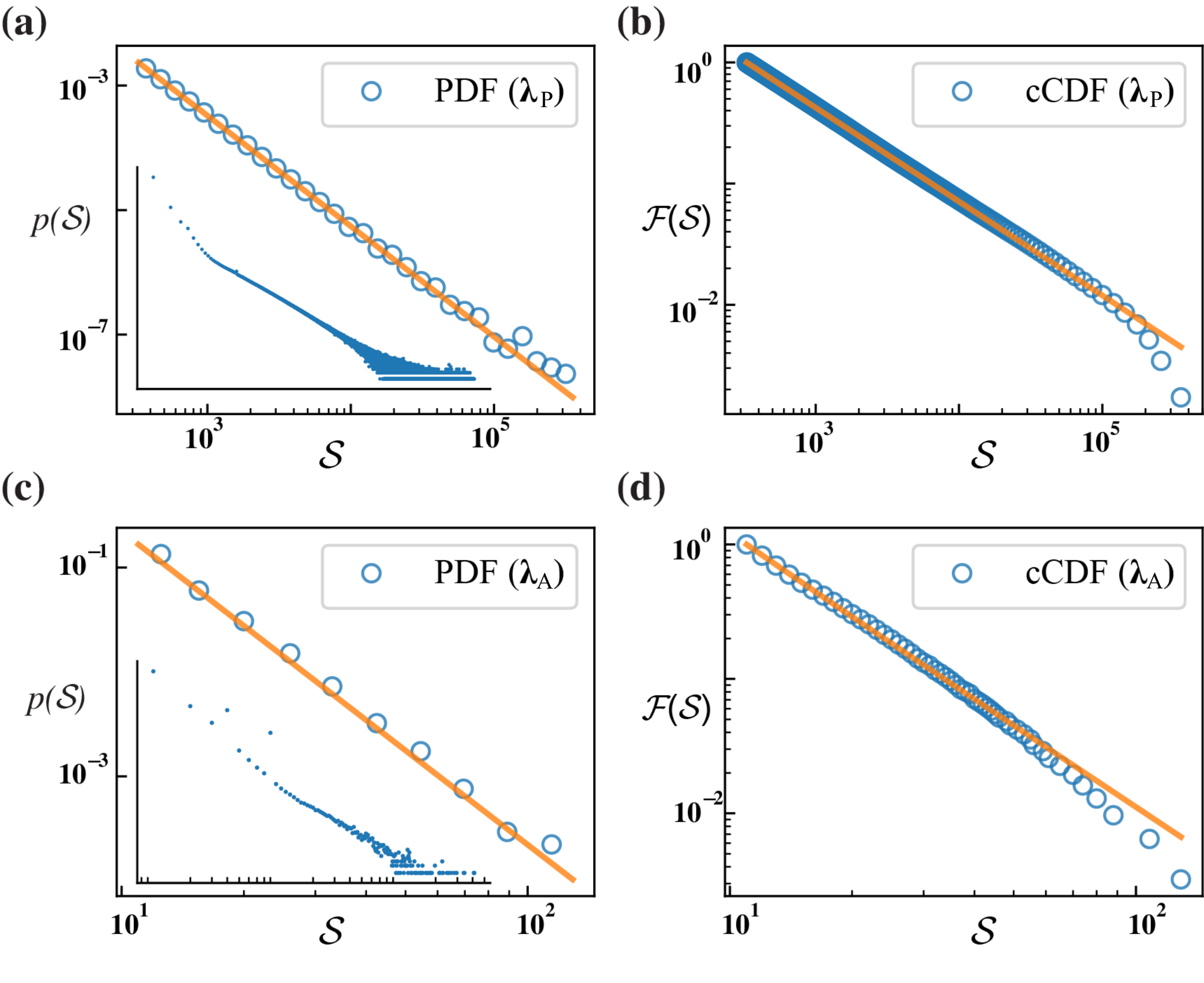}
\caption{\textbf{The results of KS method at two critical points.} \textbf{(a-c)} show PDFs with logarithmic-binning and power-law fits by KS methods;
the insets show the original PDFs (linear-binning and no truncation). \textbf{(b-d)} show the corresponding log-binned cCDFs with power-law fits. \textbf{(a-b)} are results at $\lP$; \textbf{(c-d)} are results at $\lA$.}
\label{fig:KSResults}
\end{figure}

Next, we define the cumulative distribution function (CDF) as the sum of probabilities up to  $p(\cS < s)$, and the complementary cumulative distribution function (cCDF) as  $\mathcal{F}(\cS) = 1 - p(\cS < s)$. To further reduce statistical fluctuations coming from each individual sample, we apply logarithmic binning, resulting in the plots shown in \MainFigref{fig:lambda_1_KS} and \MainFigref{fig:lambda_2_KS}.

Previous studies have demonstrated that applying the Kolmogorov-Smirnov (KS) method to the log-binned cCDF yields more reliable results compared to applying it directly to the raw PDF \cite{Clauset2009}. This is because the cCDF and log-binning smooth out the statistical fluctuations inherent in raw data, providing a more stable statistical measure. To ensure the robustness of our results, we follow the same methodology here. \SupNoteFigref{fig:KSResults} demonstrates log-binned PDF and cCDFs for $\lP$ and $\lA$. The difference between logarithmic-binning and the conventional linear-binning is that logarithmic-binning divides the data into bins whose widths increase exponentially, which is useful for analyzing data that spans several orders of magnitude. This approach ensures that each bin contains a sufficient number of data points even in the tails of the distribution, thereby reducing noise and providing a clearer representation of the underlying distribution. In contrast, linear-binning divides the data into equally spaced bins, which can lead to sparsity and high statistical fluctuations in regions where data points are scarce, especially when dealing with heavy-tailed distributions. Note that since we set the bins' interval on log-scaled axis is the same for all plots, distributions with larger domain will have more bins -- e.g., \SupNoteFigref{fig:KSResults}b has larger domain $\sim(10^2,10^6)$ compared to \SupNoteFigref{fig:KSResults}d whose domain $\sim(10^1,10^3)$, thus \SupNoteFigref{fig:KSResults}b has denser data points.

\begin{figure}[ht!]
\includegraphics[width=0.95\columnwidth]{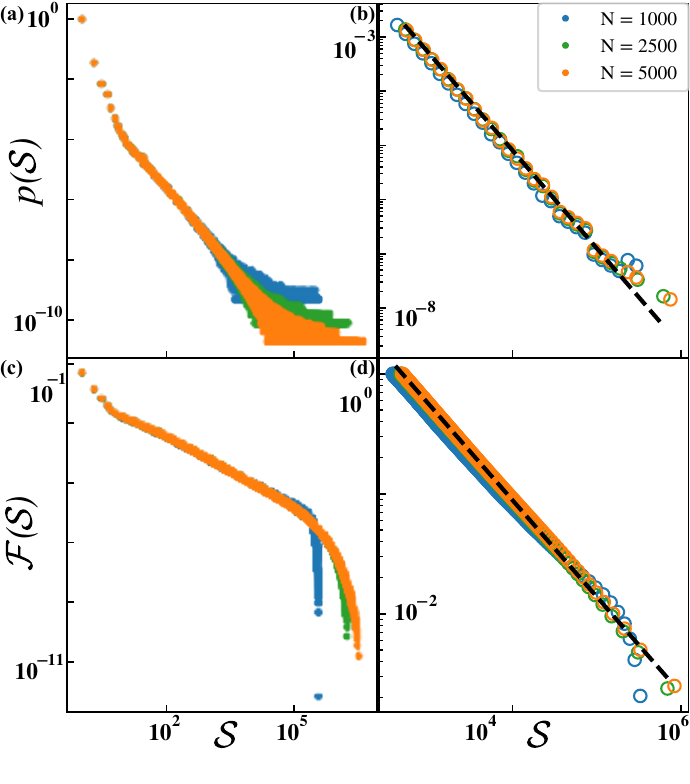 }
\caption{\textbf{Grid size invariance of cluster distributions at critical threshold $\lP$}. \textbf{(a)} shows how the PDFs evolve with increasing grid size ($N$). As the grid size increases, the tail extends further. \textbf{(b)} shows the power law fits to log-binned PDFs. It can be seen that, as the grid size evolves, the PDF behavior remains consistent. \textbf{(c)} shows how the cCDFs evolve with increasing $N$. The tail extension is directly apparent, and the exponential cutoff term depends on the value of $N$. \textbf{(d)} shows the power law fits to log-binned cCDFs. It can be seen that the consistent power-law behavior is preserved across increasing grid sizes, and the fitted exponents overlap. The black dashed lines represent the power-law fits, and for all three grid sizes, the fits overlap remarkably well. 
We note here that for small grids where finite-size effects are apparent, KS method would decide on power-law with exponential cut-off; while for large grids, the power-law with exponential cut-off is no longer decisively favored over the plain power-law, and in the thermodynamic limit ($N\to\infty$), the distribution approaches a pure power-law.}
\label{fig:SizeEvolutionofDists}
\end{figure}
\subsection{Size Evolution of Cluster Distributions at $\lA$}\label{sec:SizeEvolution}

For $\lP$, \SupNoteFigref{fig:SizeEvolutionofDists} shows that the cluster distribution extends up to the system size regardless of grid size ($N$). The exponential cutoff appears only due to the system's finite size, with the cutoff point shifting to larger values as $N$ increases. This analysis was not performed for $\lA$ since criticality at $\lA$ is not constrained by grid size and its distribution does not extend to the system size. Consistent power-law behavior for $\lP$ is observed, as indicated by the overlapping on black dashed fit, confirming the presence of percolation behavior. In the power-law relation presented in Eq.~\ref{lambda_1_fit_results}, we use a grid size of \( N = 5000 \).

\section{Kolmogorov-Smirnov Method}\label{sec:KS method}

One important fact about empirical power-law data is that the scaling is rarely valid for the full range of the data. More often, the power law applies only for values greater than some minimum $\cS_\mrm{min}$, i.e., only the tail follows a power law. Kolmogorov-Smirnov (KS) method~\cite{Clauset2009, alstott_powerlaw_2014, barabasi2013network} is proposed to determine the $\tau$ and $\cS_\mrm{min}$, test the goodness-of-fit, and compare between alternative fat-tailed models via Log-likelihood ratio test in a principled manner.

The optimal $\cS_\mrm{min}$ minimizes the relative KS statistic between the empirical data and the fitted model while the optimal $\tau$ maximizes the likelihood of the data given the model. However, fitting the data and obtaining ($\tau$, $\cS_\mrm{min}$) alone does not tell us how well the power-law model fits the data; thus, we need a goodness-of-fit test that returns a $p$-value quantifying the plausibility of the power law hypothesis ($p_\mathrm{gf}$). The closer $p_\mathrm{gf}$ is to 1, the more likely it is that the difference between the empirical data and the model can be attributed to statistical fluctuations alone. If $p_\mathrm{gf}$ is very small, the model is not a plausible fit to the data. \citet[Chap. 4]{barabasi2013network} suggests the model is accepted if $p_\mathrm{gf} > 0.01$, while \citet{Clauset2009} proposes a harsher threshold of $p_\mathrm{gf} > 0.1$. We adopt the latter.

Even if we obtain a plausible power-law fit, it does not guarantee that the power law ($\propto x^{-\tau}$) is the best model. To rigorously assess its suitability, we must compare the power-law model against alternative fat-tailed distributions. Following the approach of \citet{Clauset2009}, we apply the KS method, including the following set of alternatives: power law with exponential cutoff ($\propto x^{-\tau}\mathrm{e}^{-\lambda x}$), exponential ($\propto \mathrm{e}^{-\lambda x}$), stretched exponential ($\propto x^{\beta-1} e^{-\lambda x^\beta}$), and log-normal ($\propto \frac{1}{x} \exp\left[ -\frac{(\ln x-\mu)^2}{2\sigma^2} \right]$).

\subsection{KS statistics $\&$ KS test}

The Kolmogorov-Smirnov statistic (KS statistic) measures the distance of two probability distributions. It's able to quantify how dissimilar the empirical distribution is from the theoretical distribution / fitted model. Utilizing KS statistic, the KS test is a nonparametric test of the equality of probability distributions that can be used to test whether a sample came from a given reference probability distribution, i.e., to test the goodness-of-fit.

Formally, for discrete data (as the cluster sizes in our case), the KS statistic is defined as the maximum distance between the cCDF of the empirical data and the cCDF of the fitted model:
\begin{equation}\label{eq:KS_statistic}
    D = \max_{\cS:\,\cS>\cS_\mrm{min}} \abs{F(\cS) - F_{\text{model}}(\cS)}
\end{equation}
Although commonly the KS statistic is defined between CDFs, it is equivalent to the above definition between cCDFs.

\subsection{Fitting procedure} 

Provided that the lower bound $\cS_\mrm{min}$ is known (the estimation of $\cS_\mrm{min}$ is discussed later), the maximum likelihood estimator (MLE) of the power-law exponent $\tau$ is given by the solution to the transcendental equation:
\begin{equation}\label{eq:alpha_MLE_exact}
    \frac{\partial_{\hat{\tau}} ~\zeta(\hat{\tau}, \cS_\mrm{min})}{\zeta(\hat{\tau}, \cS_\mrm{min})} = -\frac{1}{n} \sum_{i=1}^{n} \ln \cS_i
\end{equation}
where $\{ \cS_i \}$ are all the observed cluster sizes $\geq \cS_\mrm{min}$. This is equivalent to maximizing the log likelihood function:
\begin{equation}\label{eq:log_likelihood}
    \mathcal{L} = -n \ln \zeta(\tau, \cS_\mrm{min}) - \tau \sum_{i=1}^{n} \ln \cS_i
\end{equation}

Though no closed-form solution exists for Supplementary~\Eref{eq:alpha_MLE_exact}, one can reliably approximate $\tau$ as:
\begin{equation}\label{eq:alpha_MLE}
    \hat{\tau} \simeq 1 + n \left[ \sum_{i=1}^{n} \ln \frac{\cS_i}{\cS_\mrm{min}-1/2} \right]^{-1}
\end{equation}

This approximation is substantially easier to compute and is accurate if $\cS_\mrm{min}$ is not too small, with error decaying fast as $\cO(\cS_\mrm{min}^{-2})$. If $\cS_\mrm{min}$ is unknown, the estimation of $\cS_\mrm{min}$ is the one minimizing the KS statistic:
\begin{align}\label{eq:s_min_KS}
    \hat{\cS}_{min} &= \argmin_{\cS'} D(\cS') \notag \\
    &= \argmin_{\cS'} \left(\max_{\cS:\,\cS>\cS'} \abs{F(\cS) - F_{\text{model}}(\cS)}\right)
\end{align}

\begin{table}[ht!]
\renewcommand{\arraystretch}{2}  
\setlength{\arrayrulewidth}{0.3mm}  
\centering
\resizebox{1\columnwidth}{!}{
\begin{tabular}{|P{4cm}|P{3cm}|P{4cm}|}  
\hline
\multirow{1}{*}{\centering \textbf{Distribution Name}} & \multicolumn{1}{c|}{\textbf{$f(x)$}} & \multicolumn{1}{c|}{\textbf{$C$}} \\
\hline
\centering Power law & $x^{-\tau}$ & $(\tau-1)x_{\mathrm{min}}^{\tau-1}$ \\
\hline
\centering Power law with cutoff & $x^{-\tau}\mathrm{e}^{-\lambda x}$ & $\frac{\lambda^{1-\tau}}{\Gamma(1-\tau,\lambda x_{\mathrm{min}})}$ \\
\hline
\centering Exponential & $\mathrm{e}^{-\lambda x}$ & $\lambda e^{\lambda x_{\mathrm{min}}}$ \\
\hline
\centering Stretched exponential & $x^{\beta-1} e^{-\lambda x^\beta}$ & $\beta\lambda e^{\lambda x_{\mathrm{min}}^\beta}$ \\
\hline
\centering Log-normal & $\frac{1}{x} \exp\left[ -\frac{(\ln x-\mu)^2}{2\sigma^2} \right]$ & $\sqrt{\frac{2}{\pi\sigma^2}}\left[ \erfc\left( \frac{\ln x_{\mathrm{min}}-\mu}{\sqrt{2}\sigma} \right) \right]^{-1}$ \\
\hline
\end{tabular}
}\caption{Definition of the power-law distribution and other statistical distributions in our reference distribution set. For each distribution, we give the kernel $f(x)$ and the normalization factor $C$ s.t. $\int_{x_{\mathrm{min}}}^\infty Cf(x)\>\mathrm{d} x=1$.}
\label{table:distributions}
\end{table}

\subsection{Goodness-of-fit test}\label{sec:goodness_of_fit}

To obtain the goodness-of-fit p-value, the commonly used procedure involves the following steps:
 
\begin{enumerate}
    \item Take the KS distance between the empirical cCDF and the best fit, denoted as $D_{\mathrm{real}}$.
    \item Plug in the best-fit parameters ($\tau$, $\cS_\mrm{min}$) into \Eref{eq:dicrete_poweR_Waw} and generate a synthetic dataset of the same size as the original dataset. Calculate the KS distance between the synthetic cCDF and the best-fit model, denoted as $D_{\mathrm{syn}}$.
    \item The goal is to see if the obtained $D_{\mathrm{syn}}$ is comparable to $D_{\mathrm{real}}$. For this, we repeat step (2.) $M$ times ($M \gg 1$, typically $10^3 \sim 10^4$), each time generating a new synthetic dataset, eventually obtaining the $p(D_{\mathrm{syn}})$ distribution. If $D_{\mathrm{real}}$ is close to the mode of $p(D_{\mathrm{syn}})$ distribution, the power law is a considered plausible. $M$.  is set to 2500 to obtain all our reported $p_\mathrm{gf}$.

    \item Assign a $p$-value ($p_{\mrm{gf}}$) to the $p(D_{\mathrm{syn}})$ distribution:
    \begin{equation}\label{eq:gof_p}
        p_{\mrm{gf}} = \int_{D}^{\infty} p(D_{\mathrm{syn}}) \, dD_{\mathrm{syn}}
    \end{equation}
    The closer $p_{\mrm{gf}}$ is to 1, the more likely it is that the difference between the empirical data and the model can be attributed to statistical fluctuations alone. If $p_{\mrm{gf}}$ is very small, the model is not a plausible fit to the data. ~\citet{barabasi2013network} suggest the model is accepted if $p_{\mrm{gf}} > 0.01$, while  ~\citet{Clauset2009} suggest a harsher threshold of $p_{\mrm{gf}} > 0.1$. We adopt the latter.
\end{enumerate}

\begin{figure}[t!]
\centering
\includegraphics[width=\linewidth]{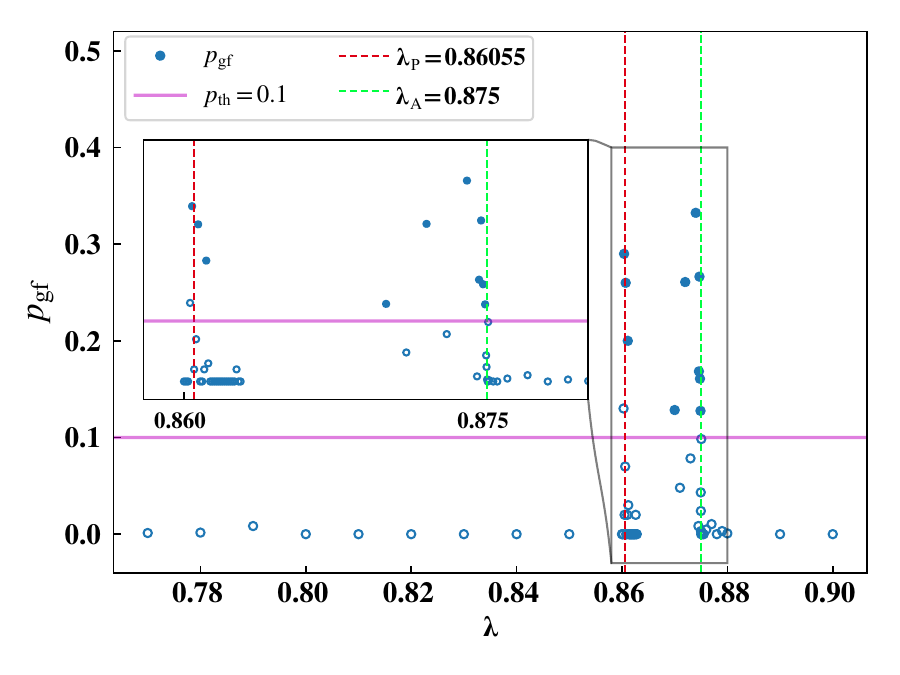}
\caption{ \textbf{Power-law fits of cluster size distributions in the logistic GOL}. The (goodness-of-fit) $p_{\mrm{gf}}$ values of the plausibility test for different $\lambda$ values. In points with $p_{\mrm{gf}}>0.1$, the hypothesis that the distribution follows a power law is favored. Note the clear peaks exceeding $0.1$ close to $\lP$ and $\lA$ in the inset. The empty circles with $p_{\mrm{gf}}>0.1$ denote points where the distribution passes the plausibility test, but fails the log-likelihood ratio test.}
\label{fig:gof_pvalues}
\end{figure}

Based on this calculation, we discuss goodness of fit results for different parameter values in the range $0.8<\lambda<0.9$. In \SupNoteFigref{fig:gof_pvalues}, we plot the plausibility test values, $p_{\mrm{gf}}$, to identify parameter ranges where the power law is a good fit for the cluster size distribution data from simulations. The peaks of high $p_{\mrm{gf}}$ values near $\lP$ and $\lA$ (filled circles in \SupNoteFigref{fig:gof_pvalues}) show that the power law is a plausible fit only near the critical points.

\begin{table}[ht!]
\renewcommand{\arraystretch}{1}  
\setlength{\arrayrulewidth}{0.3mm}  
\centering
\resizebox{1\columnwidth}{!}{
\begin{tabular}{|P{3cm}|P{1cm}@{\hskip 1cm}P{1cm}|P{1 cm}@{\hskip 1cm}P{1 cm}|}  
\hline
\multirow{2}{=}{\centering Alternatives to \\ power law ($ x^{-\tau} $) distribution } & \multicolumn{2}{c|}{\rule{0pt}{1.2\normalbaselineskip} $\lP = 0.86055$} & \multicolumn{2}{c|}{\rule{0pt}{1.2\normalbaselineskip} $\lA = 0.875$} \\

\cline{2-5}
 & \rule{0pt}{1.4\normalbaselineskip} LR & \rule{0pt}{1.4\normalbaselineskip} \textbf{$p_\mathrm{LR}$} & \rule{0pt}{1.4\normalbaselineskip} LR & \rule{0pt}{1.4\normalbaselineskip} \textbf{$p_\mathrm{LR}$} \\
 
\hline
\centering \rule{0pt}{1.2\normalbaselineskip} Log-Normal & 
\raisebox{-2.5ex}{-0.189} & \raisebox{-2.5ex}{0.69} & 
\raisebox{-2.5ex}{-0.41} & \raisebox{-2.5ex}{0.54} \\
\centering ($\frac{1}{x} \exp\left[ -\frac{(\ln x-\mu)^2}{2\sigma^2} \right]$) & & & & \\
\hline
\centering \rule{0pt}{1.2\normalbaselineskip}Stretched exponential & 
\raisebox{-3.3ex}{-0.97} & \raisebox{-3.3ex}{0.61} & 
\raisebox{-3.3ex}{-0.31} & \raisebox{-3.3ex}{0.80} \\
\centering ($ x^{\beta-1} e^{-\lambda x^\beta}$) & & & & \\
\hline
\centering \rule{0pt}{1.2\normalbaselineskip} Exponential & 
\raisebox{-2.5ex}{373} & \raisebox{-2.5ex}{\textbf{0.001}} & 
\raisebox{-2.5ex}{27.6} & \raisebox{-2.5ex}{\textbf{0.005}} \\
\centering  \raisebox{0ex}{($ \mathrm{e}^{-\lambda x}$)} & & & & \\
\hline
\centering \rule{0pt}{1.2\normalbaselineskip}Power law with cutoff & 
\raisebox{-3ex}{-3.89} & \raisebox{-3ex}{\textbf{0.005}} & 
\raisebox{-3ex}{-0.84} & \raisebox{-3ex}{0.70} \\
\centering ($ x^{-\tau}\mathrm{e}^{-\lambda x}$) & & & & \\
\hline
\multirow{3}{*}{\centering \rule{0pt}{2\normalbaselineskip}Verdict} & \multicolumn{2}{c|}{Good support for} & \multicolumn{2}{c|}{Good support for} \\
 & \multicolumn{2}{c|}{power law with cutoff} & \multicolumn{2}{c|}{power law} \\
 & \multicolumn{2}{c|}{$p_\mathrm{gf}=\textbf{0.46}$} 
 & \multicolumn{2}{c|}{$p_\mathrm{gf}=\textbf{0.13}$} \\
\hline
\end{tabular}
}
\caption{The plausibility $p_\mathrm{gf}$-values (goodness-of-fit test) for power-law and log-likelihood ratio test results between the power-law and alternative distributions at two critical points. Statistically significant $p$-values are denoted in \textbf{bold}. The plausibility values both exceed 0.1, meaning the power-law is a plausible fit for both cases. $LR$ is the log-likelihood ratio of power-law against alternative distributions: power-law with exponential cutoff ($\propto x^{-\tau}\mathrm{e}^{-\lambda x}$), exponential ($\propto \mathrm{e}^{-\lambda x}$), stretched exponential ($\propto x^{\beta-1} e^{-\lambda x^\beta}$), and log-normal ($\propto \frac{1}{x} \exp\left[ -\frac{(\ln x-\mu)^2}{2\sigma^2} \right]$). If $LR>0$, the power-law model is favored; if $LR<0$, the alternative distribution is favored. The $p_\mathrm{LR}$-value of log-likelihood ratio test denotes the significance of the sign of $LR$: if $p_\mathrm{LR}<0.05$, the sign of $LR$ is considered significant. The ones at $\lP$ and $\lA$ indicate that power-law distribution is favored over exponential distribution. The other one at $\lP$ shows that power-law with exponential cutoff is favored over power-law. The final column lists the final judgments of the statistical support for the power-law hypothesis at each critical point: ``with cutoff'' means that the conclusion is power-law with exponential cutoff, while ``good'' indicates that the power-law is a good fit and none of the alternatives considered is favored. 
Note that this table is reported on a grid where finite-size effect is consequential; for a much larger grid, at $\lP$ the $p_{LR}$ of power law v.s. power law with cutoff would be $>0.05$, i.e. insignificant.}
\label{table:statconclusion}
\end{table}

\subsection{Model comparison and statistical results}\label{model:goodness}

Even if we obtain a plausible power-law fit, it does not guarantee that the power law ($\propto x^{-\tau}$) is the best model. To rigorously assess its suitability, we must compare the power-law model against alternative fat-tailed distributions. Following the approach of \citet{Clauset2009}, we apply the KS method, including the following set of alternatives: power law with exponential cutoff, exponential, stretched exponential, and log-normal. The definition of these distributions is given in \SupNoteTableref{table:distributions}.

A common method to compare models is the likelihood ratio test -- to compute the likelihood of the data under two competing distributions, and take the logarithm of the ratio of the two likelihoods, denoted by $\mathcal{LR}$. \\

\begin{gather}
    \mathcal{R} = \frac{\mathcal{L}_1}{\mathcal{L}_2} = \prod_{i=1}^{n} \frac{p_1(\cS_i)}{p_2(\cS_i)} \notag \\
    \mathcal{LR} = \ln \mathcal{R} = \ln \mathcal{L}_1 - \ln \mathcal{L}_2
\end{gather}

If $\mathcal{LR}$ is positive, the first distribution is favored; if negative, the second distribution is favored; if close to zero, the data are insufficient to favor either model. We further apply the method proposed by Vuong \cite{vuong1989likelihood} which gives a $p$-value ($p_\mathrm{LR}$) that tells us whether the observed sign of $\mathcal{LR}$ is statistically significant. If this $p_\mathrm{LR}$-value is small (typically, $p_\mathrm{LR} < 0.05$), then the sign is a reliable indicator of which model is a better fit to the data.

\SupNoteTableref{table:statconclusion} presents the results of the goodness-of-fit and log-likelihood ratio tests, based on a sample size of $N=5000$. Statistically significant $p_\mathrm{LR}$-values are denoted in \textbf{bold}. Note that for goodness-of-fit test results, the larger the $p_\mathrm{gf}$ value, the more plausible the power-law model is. Whereas for log-likelihood ratio test results, the larger the $p_\mathrm{LR}$-value, the less significant the sign of the test is. The final column lists our judgment of the statistical support for the power-law hypothesis at each critical point.

\begin{figure}[t!]
\centering
\includegraphics[width=\linewidth]{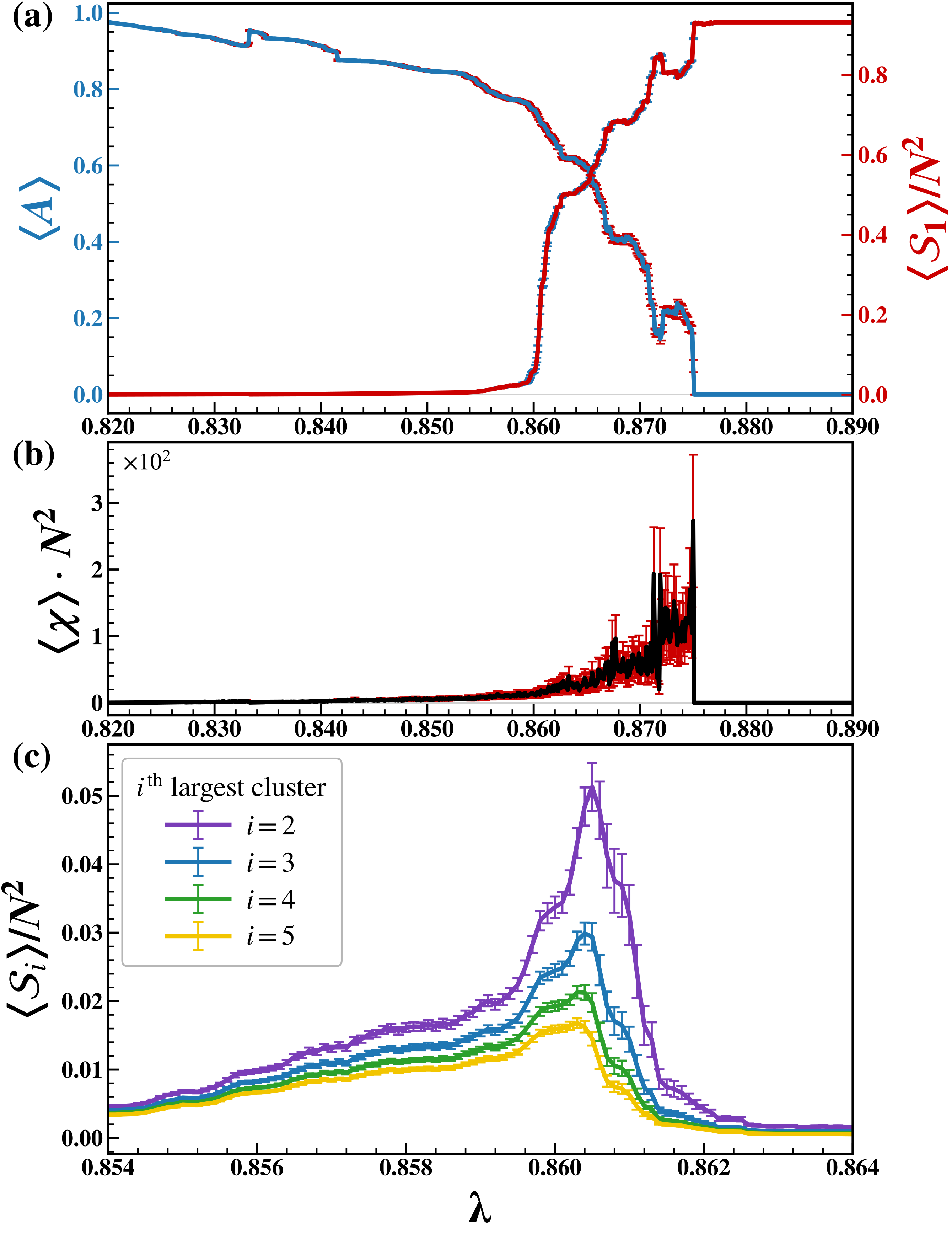}
\caption{\textbf{Standard-error comparison of key observables and subleading cluster sizes.}
\textbf{(a)} Asymptotic mean activity $\langle A\rangle$ (blue curve; red error bars) and normalized largest-cluster size $\langle \mathcal{S}_1\rangle/N^2$ (red curve; blue error bars) both exhibit very small standard errors of their sample means over independent realizations at each $\lambda$ (error bars), indicating stable estimates with minimal sampling variability.
\textbf{(b)} Susceptibility $\langle \chi\rangle \cdot N^{2}$ (black curve with black error bars) shows markedly larger standard errors near the peak, reflecting enhanced fluctuations close to the transition.
\textbf{(c)} Average normalized largest ranking cluster sizes $\langle \mathcal{S}_i \rangle / N^2$ shows standard errors remain moderate over most of the range but increase near the critical point due to the scale-invariant nature of $\lP$.}
\label{fig:activity_chi_standarderror}
\end{figure}

\section{Standard-error Analysis for $\langle A\rangle$, $\langle \chi\rangle$, and $\langle \mathcal{S}_i\rangle$}\label{sec:standarderror}

To evaluate the statistical reliability of the computed observables, we analyze the standard errors associated with $\langle A\rangle$, $\langle \chi\rangle$, and $\langle \mathcal{S}_i\rangle$. As shown in \SupNoteFigref{fig:activity_chi_standarderror} (a-b), the asymptotic mean activity $\langle A\rangle$ and the normalized largest-cluster size $\langle \mathcal{S}_1\rangle/N^2$ exhibit very small standard errors across the range of $\lambda$, confirming the stability and consistency of their analysis across different realizations. In contrast, the susceptibility $\langle \chi\rangle/N^2$ shows markedly larger standard errors near $\lA$, due to enhanced fluctuations in activity close to the transition point.

We perform a similar analysis for the subleading clusters in \SupNoteFigref{fig:activity_chi_standarderror} (c), which shows consistent behavior across $\langle \mathcal{S}_i\rangle / N^2$ for $i=2$–$5$. The average normalized sizes $\langle \mathcal{S}_i\rangle / N^2$ for the $i^{\text{th}}$ largest clusters display moderate standard errors that increase near the critical point $\lP$. This behavior arises from the scale-invariant nature of $\lP$ and can be understood in terms of the high variance inherent in the power-law cluster-size distribution near criticality. Since $\mathcal{S}_i$ values are sampled from this heavy-tailed distribution around $\lP$, they exhibit higher variability, leading to larger standard errors in $\langle \mathcal{S}_i\rangle$ near the transition.

\begin{figure}[t!]
\centering
\includegraphics[width=\linewidth]{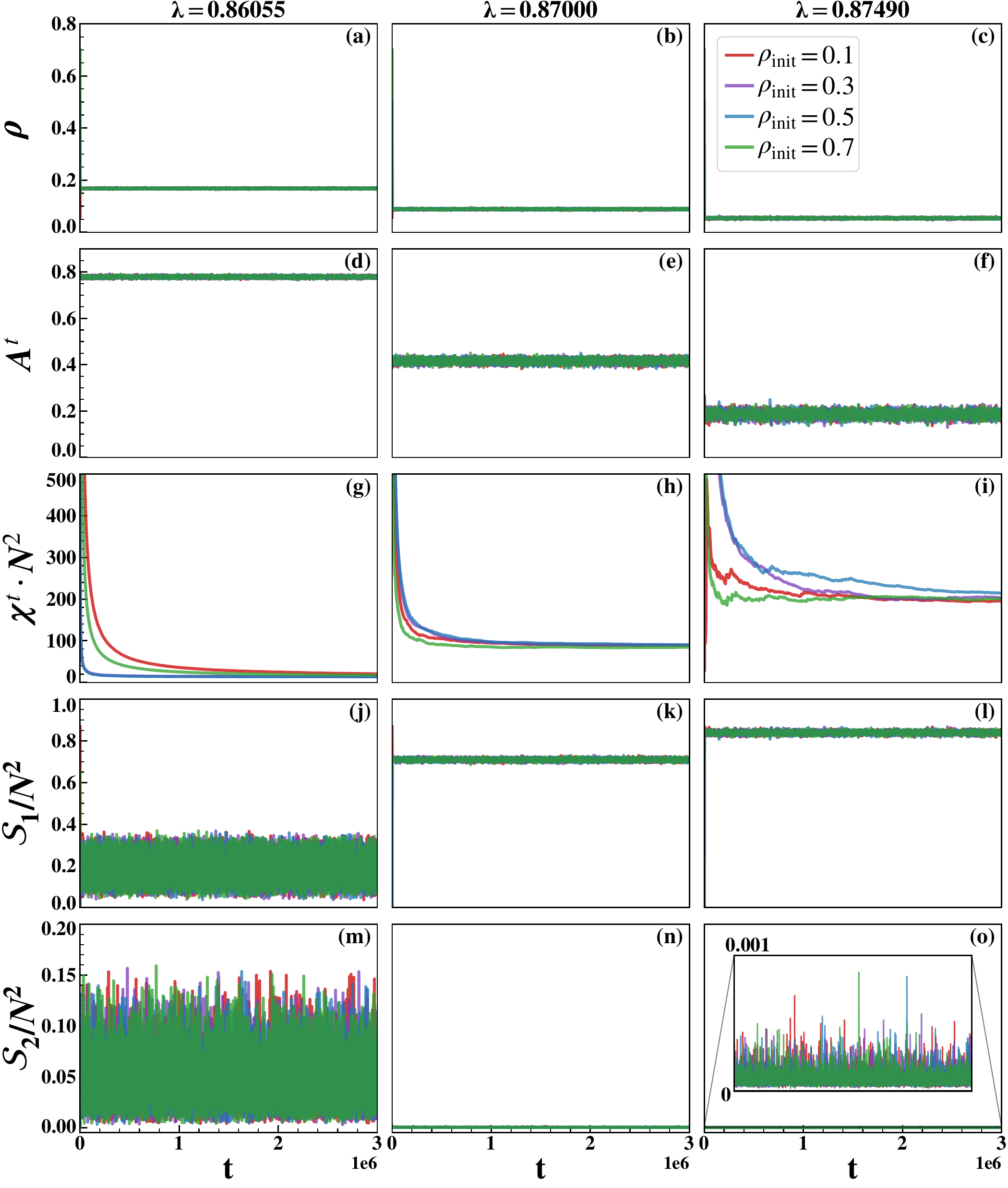}
\caption{\textbf{Role of random initial conditions.} For random initial conditions with densities $p_{\text{init}}\in\{0.1,0.3,0.5,0.7\}$ and representative values of the control parameter $\lambda\in\{0.8655,0.87,0.8749\}$, we show the time evolution of the density $p$ \textbf{(a–c)}, activity $A^t$ \textbf{(d–f)}, susceptibility $\chi^t N^2$ \textbf{(g–i)}, largest cluster size fraction $\mathcal{S}_1/N^2$ \textbf{(j–l)}, and second largest cluster size fraction $\mathcal{S}_2/N^2$ \textbf{(m–o)}. In all cases, the long-time thermodynamic behavior converges to values set by the deterministic parameter $\lambda$, demonstrating that the stationary statistical properties are governed by $\lambda$ and are essentially independent of the random initial state. As $\lambda$ increases from $\lP$ to $\lA$, the asymptotic activity $A^t$ decreases while its temporal fluctuations grow, as reflected in the increase of $\chi^t N^2$. At the same time, $\mathcal{S}_1/N^2$ increases whereas $\mathcal{S}_2/N^2$ decreases, with the largest fluctuations of both occurring near $\lP$, where clusters of different rank compete and are most self-similar. Lastly, \textbf{(o)} illustrates via a zoomed-in inset that, even beyond $\lP$, a subleading second largest cluster persists and fluctuates on shorter length scales.
}  
\label{fig:3panel}
\end{figure} 

\vspace{1cm} 
\section{Asymptotic Dynamics Independent of Initial Randomness}\label{sec:invariancetoinitialconds}
Stochasticity enters only through the construction of the initial configuration; once this configuration is fixed, the subsequent time evolution is uniquely determined by the control parameter $\lambda$. For all different random initializations, the system relaxes to a common stationary statistical state that depends only on $\lambda$ and is independent of both the initialization density and the microscopic details of the random initial configuration.  

As shown in \SupNoteFigref{fig:3panel}, for initialization densities $p_{\mathrm{init}} \in \{0.1, 0.3, 0.5, 0.7\}$, the long-time trajectories of $p$, $A$, $\chi$, $\mathcal{S}_1$, and $\mathcal{S}_2$ approach common asymptotic values. This demonstrates that stochastic elements in the initialization play no role in the steady-state observables: for each fixed $\lambda$, the dynamics converge to a well-defined $\lambda$–dependent statistical attractor in configuration space. Therefore, in contrast to models with deterministic update rules but random control parameters, our system is controlled by a single deterministic parameter $\lambda$, which fully specifies the statistical behavior.

More specifically, the observed behaviors are consistent with our previous discussion. \SupNoteFigref{fig:3panel} (a–f) shows that the overall density $p$ and activity $A^t$ decrease with increasing $\lambda$, due to the growing dominance of the largest $0$-state cluster in the grid, as seen in \SupNoteFigref{fig:3panel} (j–l). At the same time, the susceptibility $\chi$ increases with $\lambda$ [\SupNoteFigref{fig:3panel} (g–i)] and reaches its maximum at $\lA$, indicating sparsely distributed, mobile active sites that are nonuniformly spread across the grid. These active sites intermittently generate and reshape $0$-state islands, leading to the fluctuations in \SupNoteFigref{fig:3panel} (o), leading to the reported power-law cluster-size distribution, which we attribute to self-organized criticality.

For $\lP$, on the other hand, the fluctuations of $\mathcal{S}_1$ and $\mathcal{S}_2$ are largest, as seen in \SupNoteFigref{fig:3panel} (j–m). Moreover, at this point the numerical values of $\mathcal{S}_1$ and $\mathcal{S}_2$ are closest to each other. Both features stem from the self-similar nature of $\lP$ and the competition between clusters of different rank. As $\lambda$ is increased beyond $\lP$, one cluster gradually dominates the grid, and the corresponding fluctuations at this length scale are reduced.

\bibliography{finite_GoL}

\end{document}